\documentclass[traditabstract]{aa} 
\usepackage{txfonts}
\usepackage{natbib}                                   
\usepackage{graphicx}
\usepackage{lscape}
\usepackage{txfonts}

\def\msun{M$_\odot$}

\begin{document}

   \title{The 2XMMi/SDSS Galaxy Cluster Survey}

   \subtitle{I. The first cluster sample and X-ray luminosity-temperature relation}

   \author{A. Takey\inst{1,2}
      \and A. Schwope\inst{1}
      \and G. Lamer\inst{1}
          }

   \institute{Leibniz-Institut f{\"u}r Astrophysik Potsdam (AIP),
              An der Sternwarte 16, D-14482 Potsdam, Germany \\
              \email{atakey@aip.de}
      \and  National Research Institute of Astronomy and Geophysics (NRIAG), 
             Helwan, Cairo, Egypt             
             }

   \date{Received .... / Accepted ....}

 
\abstract 
{We present a catalogue of X-ray selected galaxy clusters and groups as 
a first release of the 2XMMi/SDSS Galaxy Cluster Survey. The survey is a search
for galaxy clusters detected serendipitously in observations with XMM-Newton
in the footprint of the Sloan Digital Sky Survey (SDSS). 
The main aims of the survey are to identify new X-ray galaxy clusters,
investigate their X-ray scaling relations, identify distant cluster candidates, 
and  study the correlation of the X-ray and optical properties. In this paper, 
we describe the basic strategy to identify and characterize the X-ray cluster
candidates that currently comprise 1180 objects selected from the second
XMM-Newton serendipitous source catalogue (2XMMi-DR3). 
Cross-correlation of the initial catalogue with recently published optically
selected SDSS galaxy cluster catalogues yields photometric redshifts for 275
objects. Of these, 182 clusters have at least one member with a spectroscopic
redshift from existing public data (SDSS-DR8). We developed an automated
method to reprocess the XMM-Newton X-ray observations, determine the optimum
source extraction radius, generate source and background spectra, and derive
the temperatures and luminosities of the optically confirmed clusters.
Here we present the X-ray properties of the first cluster sample, which
comprises 175 clusters, among which 139 objects are new X-ray discoveries while
the others were previously known as X-ray sources. 
For each cluster, the catalogue provides: two identifiers, coordinates,
temperature,  flux [0.5-2] keV, luminosity [0.5-2] keV extracted from an
optimum aperture, bolometric luminosity $L_{500}$, total mass $M_{500}$,
radius $R_{500}$, and the optical properties of the counterpart.  The 
first cluster sample from the survey covers a wide range of redshifts from 0.09 to 0.61,
bolometric luminosities $L_{500} = 1.9 \times 10^{42} - 1.2 \times
10^{45}$\,erg s$^{-1}$, and masses $M_{500} = 2.3 \times 10^{13} - 4.9 \times
10^{14}$\,\msun.  We extend the relation between the X-ray bolometric
luminosity $L_{500}$ and the X-ray temperature towards significantly lower 
$T$ and $L$ and still find that the slope of the linear $L-T$ relation is  
consistent with values published for high luminosities.}


\maketitle


\section{Introduction}
Galaxy clusters are the most visible tracers of large-scale structure. They
occupy very massive dark matter halos and are observationally  accessible by
a wide range of means. Their locations are found to corresponding to large numbers of
tightly clustered galaxies, pools of hot X-ray  emitting gas, and relatively
strong features in the gravitational lensing shear field. Precise observations
of large numbers of clusters  provide an important tool for testing our
understanding of cosmology and structure formation. Clusters are also
interesting laboratories  for the study of galaxy evolution under the
influence of extreme environments \citep{Koester07}. 

The baryonic matter of the clusters is found in two forms: first, individual
galaxies within the cluster, which are most effectively studied through  optical and NIR
photometric and spectroscopic surveys; and second, a hot, ionized intra-cluster
medium (ICM), which can be studied by means of its X-ray emission and the Sunyaev-Zeldovich
(SZ) effect \citep[][]{Sunyaev72,Sunyaev80}. The detection of clusters using SZ effect is 
a fairly new and highly promising  technique for which tremendous progress has been made 
in finding high redshift clusters and measuring the total cluster mass  
\citep[e.g.][]{Planck11,Vanderlinde10,Marriage10}.   

The X-ray selection of clusters has several advantages for cosmological surveys: the
observable X-ray luminosity and temperature  of a cluster is tightly
correlated with its total mass, which is indeed its most fundamental parameter
\citep{Reiprich02}. These relations provide the ability to measure both the mass
function \citep{Boehringer02} and power spectrum \citep{Schuecker03}, which directly 
probe the cosmological models.  Since the cluster X-ray emission is
strongly peaked on the dense cluster core, X-ray selection is less affected by projection 
effects than optical surveys and clusters can be identified efficiently over
a wide redshift range.  

Many clusters have been found in X-ray observations with Uhuru, HEAO-1, Ariel-V,
Einstein, and EXOSAT, which have allowed a more accurate characterization of  their
physical proprieties (for a review, see \cite{Rosati02}). The ROSAT All Sky
Survey \citep[RASS,][]{Voges99} and the deep pointed observations  have led to
the discovery of hundreds of clusters. In ROSAT observations, 1743 clusters have been 
identified, which are compiled in a meta-catalogue called MCXC by
\cite{Piffaretti10}. The MCXC catalogue is based on published  RASS-based  
(NORAS, REFLEEX, BCS, SGP, NEP, MACS, and CIZA) and
serendipitous (160D, 400D, SHARC, WARPS, and EMSS) cluster catalogues.  

The current generation of X-ray satellites XMM-Newton, Chandra, and Suzaku have
provided follow-up observations of statistical samples of ROSAT  clusters for
cosmological studies \citep{Vikhlinin09} and detailed information on the
structural proprieties of the cluster population
\citep[e.g.][]{Vikhlinin06,Pratt10,Arnaud10}. Several projects are ongoing to
detect new clusters of galaxies from XMM-Newton and Chandra observations
(e.g. the XSC \citep{Romer01}, XDCP \citep{Fassbender07}, XMM-LSS
\citep{Pierre06}, COSMOS \citep{Finoguenov07}, SXDS \citep{Finoguenov10}, and
ChaMP \citep{Barkhouse06}). 

In this paper, we present the 2XMMi/SDSS galaxy cluster survey, a search
for galaxy clusters based on extended sources in the 2XMMi catalogue
\citep{Watson09}  in the field of view the Sloan Digital Sky Survey (SDSS). The
main aim of the survey is to build a large catalogue of new X-ray clusters  
in the sky coverage of SDSS. The catalogue will allow us to investigate the
correlation between the X-ray and optical properties of the clusters. 
One of the long term goals of the project is to improve the X-ray scaling
relations, and to prepare for the eROSITA cluster surveys,  a mid term goal is
the selection of the cluster candidates beyond the SDSS-limit for studies of
the distant universe. Here we present a first cluster sample of the survey which
comprises 175 clusters found by cross-matching the 2XMMi sample with published
SDSS based optical cluster catalogues. 

The paper is organized as follows: in Sect.~2 we describe the procedure of
selecting the X-ray cluster candidates as well as  their possible
counterparts in SDSS data. In Sect.~3 we describe the X-ray data the reduction
and analysis of the optically confirmed clusters. The discussion of the
results is described in Sect.~4. Section 5 concludes the paper. The cosmological
parameters  $\Omega_{\rm M}=0.3$, $\Omega_{\Lambda}=0.7$ and
$H_0=70$\ km\ s$^{-1}$\ Mpc$^{-1}$ were used throughout this paper.


\section{Sample construction}
We describe our basic strategy for identifying clusters among the 
extended X-ray sources in the 2XMMi catalogue. We then proceed
by cross-matching the initial catalogue with those of optically
selected galaxy clusters from the SDSS thus deriving a catalogue of X-ray
selected and optically confirmed clusters with measured redshifts, whose X-ray
properties are analysed in Sect.~3.

\subsection{X-ray cluster candidate list}
X-ray observations provide a robust method for the initial identification of
galaxy clusters as extended X-ray sources. A strategy to create a clean galaxy
cluster sample is to construct a catalogue of  X-ray cluster candidates
followed by optical observations. XMM-Newton archival observations provide the
basis for creating catalogues of serendipitously identified point-like and extended
X-ray sources. The largest X-ray source  catalogue ever produced is the 
second XMM-Newton source catalogue \citep{Watson09}.  The latest edition of
this catalogue is 2XMMi-DR3, which was released on 2010 April 28.  
The 2XMMi-DR3 covers 504 deg$^{2}$ and contains $\sim 3$ times as many discrete
sources as either the ROSAT survey or pointed catalogues. The catalogue
contains 353191 X-ray source detections corresponding to 262902 unique X-ray
sources detected in 4953 XMM-Newton EPIC (European Photon Imaging Camera)
observations made between 2000 February 3 and 2009 October 8.  

The 2XMMi-DR3 contains 30470 extended source detections, which form the primary
database for our study. This initial sample contains a very significant number
 of spurious detections caused by the clustering of unresolved point sources, edge
effects, the shape of the PSF (point spread function) of the X-ray mirrors,
large extended sources consisting of several minor sources, and other
effects such as X-ray ghosts and similar. 

We applied several selection steps to obtain a number of X-ray extended
sources that were then visually inspected individually. In our study, we considered only
sources at high Galactic latitudes, $|b| > 20\degr$, and discarded those 
that were flagged as spurious in the 2XMMi-DR3 catalogue by the screeners of
the XMM-Newton SSC (Survey Science Centre). The source detection pipeline
used for the creation of the 2XMMi catalogues allows for a maximum core radius
of extended sources of 80 arcsec. Sources with extent parameters equalling
that boundary were discarded, screening of a few examples shows that those
sources are spurious or large extended sources (the targets of the
observations) that were discarded anyhow. This initial selection reduced the
number to 4027 detections. 

Since our main aim is the generation of a serendipitous cluster sample, we
removed sources that were the targets of the XMM-Newton observation. We also
discarded fields containing large extended sources and selected only those
fields within the footprint of SDSS, which left 1818 detections.  After
removing multiple detections of the same extended sources (phenomena caused by either
problematic source geometries in a single XMM-Newton observation or 
duplicate detections in a re-observed field), the catalogue was reduced to
1520 extended sources that were regarded as unique.

This list still contains spurious detections for a number of reasons:
(a) point-source confusion, (b) resolution of one asymmetric extended source
into several symmetric extended sources, (c) the ill-known shape of the PSF
leads to an excess of sources near bright point-sources, for both 
point-like and extended sources, and (d) edge effects/low exposure times.
To remove the obvious spurious cases, we visually inspected the X-ray images 
of the initial 1520 detections using the FLIX upper limit 
server\footnote{http://www.ledas.ac.uk/flix/flix.html}. As a result, we were 
left with 1240 confirmed extended X-ray sources. 

We then made use of the SDSS to remove additional non-cluster sources. 
We downloaded the XMM-Newton EPIC X-ray images from the XMM-Newton Science
Archive \citep[XSA:][]{Arviset02} and created summed EPIC (PN+MOS1+MOS2)
images in the energy band $0.2 - 4.5$\,keV. Using these, we created smoothed
X-ray contours, which were overlaid onto co-added $r$, $i$, and $z-$band SDSS
images. Visual inspection of those optical multi-colour images with X-ray
contours overlaid, allowed us to remove extended sources corresponding to
nearby field galaxies, as well as those objects that are likely spurious 
detections. The resulting list which passes these selection criteria 
contains 1180 cluster candidates, about 75 percent of which  
are newly discovered.

\begin{figure}
  \resizebox{\hsize}{!}{\includegraphics{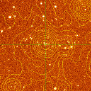}}
  \caption{The X-ray-optical overlay of the representative cluster 2XMM
    J104421.8+213029 at photometric redshift = 0.4975. The X-ray contours are
    overlaid on the SDSS co-added image obtained in  $r$, $i$, and
    $z$-bands. The field of view is $4'\times 4'$ centred on the X-ray
    cluster position. The cross-hair indicates the cluster stellar mass centre 
    as given by \cite{Szabo11}.}
    \label{f:overlay}
\end{figure}

Figure~\ref{f:overlay} shows the X-ray-optical overlay of a new X-ray 
cluster, which has a counterpart in SDSS at photometric redshift = 0.4975 and
a stellar mass centre indicated by the cross-hair as given by \cite{Szabo11}
(see Section 2.2). We use this cluster to illustrate the main steps of our
analysis in the following sections. In Appendix A, Figures A.1 to A.4 show the 
X-ray-optical overlays and the extracted  X-ray spectra for four clusters 
illustrating results for various redshifts covered by our sample and different 
X-ray fluxes. 

About one quarter of the X-ray selected cluster candidates have no plausible
optical counterpart. These are regarded as high-redshift candidates 
beyond the SDSS limit at $z \geq 0.6$, and suitable targets for dedicated
optical/near-infrared follow-up observations \citep[see e.g.~][]{Lamer08}.


\subsection{The cross-matching with optical cluster catalogues}
The SDSS offers the opportunity to produce large galaxy-cluster
catalogues. Several techniques were applied to identify likely clusters from
multiband imaging and SDSS spectroscopy. We use those published catalogues to
cross-identify common sources in our X-ray selected and those optical
samples. All these optical catalogues give redshift information per cluster,
which we use in the following to study the X-ray properties of our sources
($z_{p}$ indicates a photometric, $z_{s}$ a spectroscopic redshift). 

Table~\ref{t:sdss} lists the main properties of the optical cluster catalogues 
that we used to confirm our X-ray selection. Below we provide a very brief 
description of each of these together with the acronym used by us:

\begin{itemize}

\item{\bf GMBCG} The {\it Gaussian Mixture Brightest Cluster
  Galaxy} catalogue  \citep{Hao10}  
  consists of more than 55,000 rich clusters across the redshift
  range $0.1 < z_{p} < 0.55$ identified in SDSS-DR7.  The galaxy clusters were 
  detected by identifying the cluster red-sequence plus a brightest cluster
  galaxy (BCG). The cross-identification of X-ray cluster candidates with the 
  GMBCG within a radius of 1 arcmin yields 136 confirmed clusters.  

\item{\bf WHL} The catalogue of Wen, Han \& Liu \citep{Wen09} consists of
  39,668 clusters of galaxies drawn from SDSS-DR6 and covers the redshift
  range $0.05 < z_{p} < 0.6$. A cluster was identified if more than eight
  member galaxies of $M_{r}\le –21$ were found within a radius of 0.5\,Mpc and
  within a photometric redshift interval $z_{p} \pm 0.04(1 + z_{p})$. We
  confirm 150 X-ray clusters by cross-matching within 1 arcmin. 

\item{\bf MaxBCG} The {\it max Brightest Cluster
  Galaxy} catalogue \citep[][]{Koester07} lists 13,823 clusters in
  the redshift range $0.1 < z_{p} < 0.3$ from SDSS-DR5. The clusters were
  identified using maxBCG red-sequence technique, which uses the clustering 
  of galaxies on the sky, in both magnitude and colour, to identify groups and 
  clusters of bright E/S0 red-sequence galaxies. The cross-match with our X-ray 
  cluster candidate list reveals 54 clusters in common within a radius of 
  one arcmin. 

\item{\bf AMF} The {\it Adaptive Matched Filter} catalogue of \cite{Szabo11}
  lists 69,173 likely galaxy clusters in the redshift range $0.045 < z_{p} <
  0.78$ extracted from SDSS-DR6 using an adaptive matched filter (AMF) cluster
  finder. The cross-match yields 127 confirmed X-ray galaxy clusters. 
\end{itemize}

In the AMF-catalogue, the cluster centre is given as the anticipated centre of
the stellar mass of the cluster, while in the other three catalogues the  cluster 
centre is the position of the brightest galaxy cluster (BCG).

Many of our X-ray selected clusters have counterparts in several optical
cluster catalogues within our chosen search radius of one arcmin. In these
cases, we use the redshift of the optical counterpart, which has minimum spatial
offset from the X-ray position. Table\ref{t:sdss} 
lists in the second to last column the number of matching X-ray sources per 
optical catalog individually and in the last column the final number after 
removal of duplicate identifications. 

The unique optically confirmed X-ray cluster sample obtained by cross-matching
with the four catalogues consists of 275 objects having at least photometric 
redshifts. 

\begin{table}
\caption{Main properties of the cluster catalogues with optically (SDSS-based)
  selected entries. The last two columns give the number of matching X-ray
  selected clusters individually and cumulatively.}
\label{t:sdss}     
\centering                                    
\begin{tabular}{c c c c c c}         
\hline                       
CLG         & Nr.       & Redshift   & SDSS      & X-ray      & Nr.CLG  \\  
catalogue   & CLG       & range      &           & CLG ($1'$) & sample \\
\hline                                  
    GMBCG   & 55,000  & 0.1 - 0.55 &  DR7      &  136  & 123 \\     
    WHL     & 39,688  & 0.05 - 0.6 &  DR6      &  150  & 72 \\
    MaxBCG  & 13,823  & 0.1 - 0.3  &  DR5      &  54   & 20 \\
    AMF     & 69,173  &0.045 - 0.78&  DR6      &  127  & 60 \\
\hline
   Total    &         &            &           &       & 275 \\     
\hline 
\end{tabular}
\end{table}

After cross-identification, we found 120 clusters of the optically confirmed cluster 
sample with a spectroscopic redshift for the brightest galaxy cluster (BCG) from the 
published optical catalogues. Since the latest data release, SDSS DR8, provides more 
spectroscopic redshifts, we searched for additional spectra of BCGs 
and other member galaxies. We ran SDSS queries searching for galaxies 
with spectroscopic redshifts $z_{s(g)}$ within 1 Mpc from the X-ray 
centre. We considered a galaxy as a member of a cluster if
$|z_{p} - z_{s(g)}| < 0.05$.  

\begin{figure}
  \resizebox{\hsize}{!}{\includegraphics{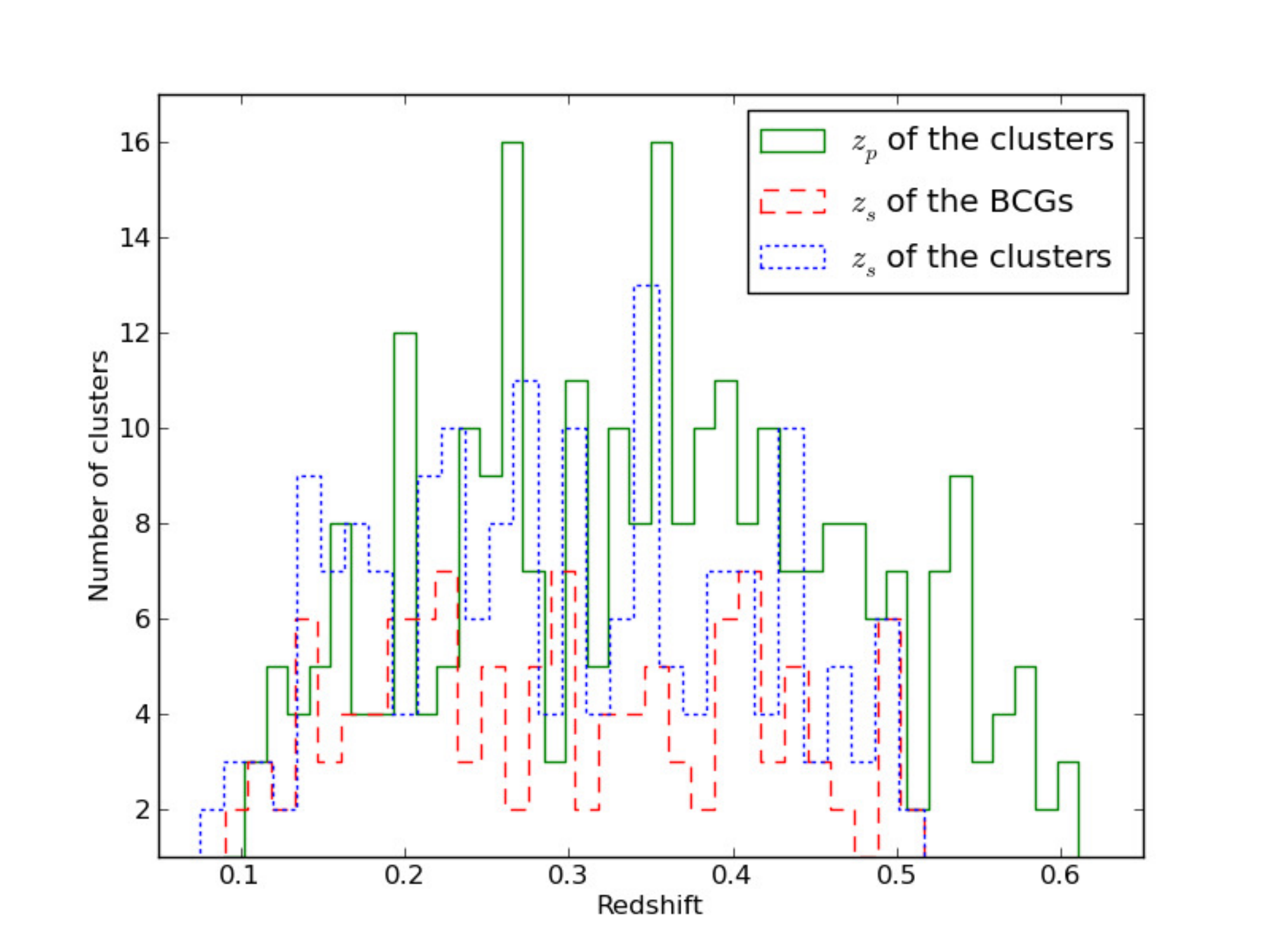}}
  \caption{The distribution of the optical redshifts for the confirmed
    clusters sample. The distribution includes the cluster photometric
    redshifts $z_{p}$ (solid line) with a median 0.36, spectroscopic redshifts
    of the BCGs $z_{s}$ (dashed line) with a median 0.3 from the optical
    cluster catalogues and the cluster  spectroscopic redshifts $z_{s}$
    (dotted line) with a median 0.3 from the SDSS data.} 
  \label{f:zdist}
\end{figure}

\begin{figure}
  \resizebox{\hsize}{!}{\includegraphics{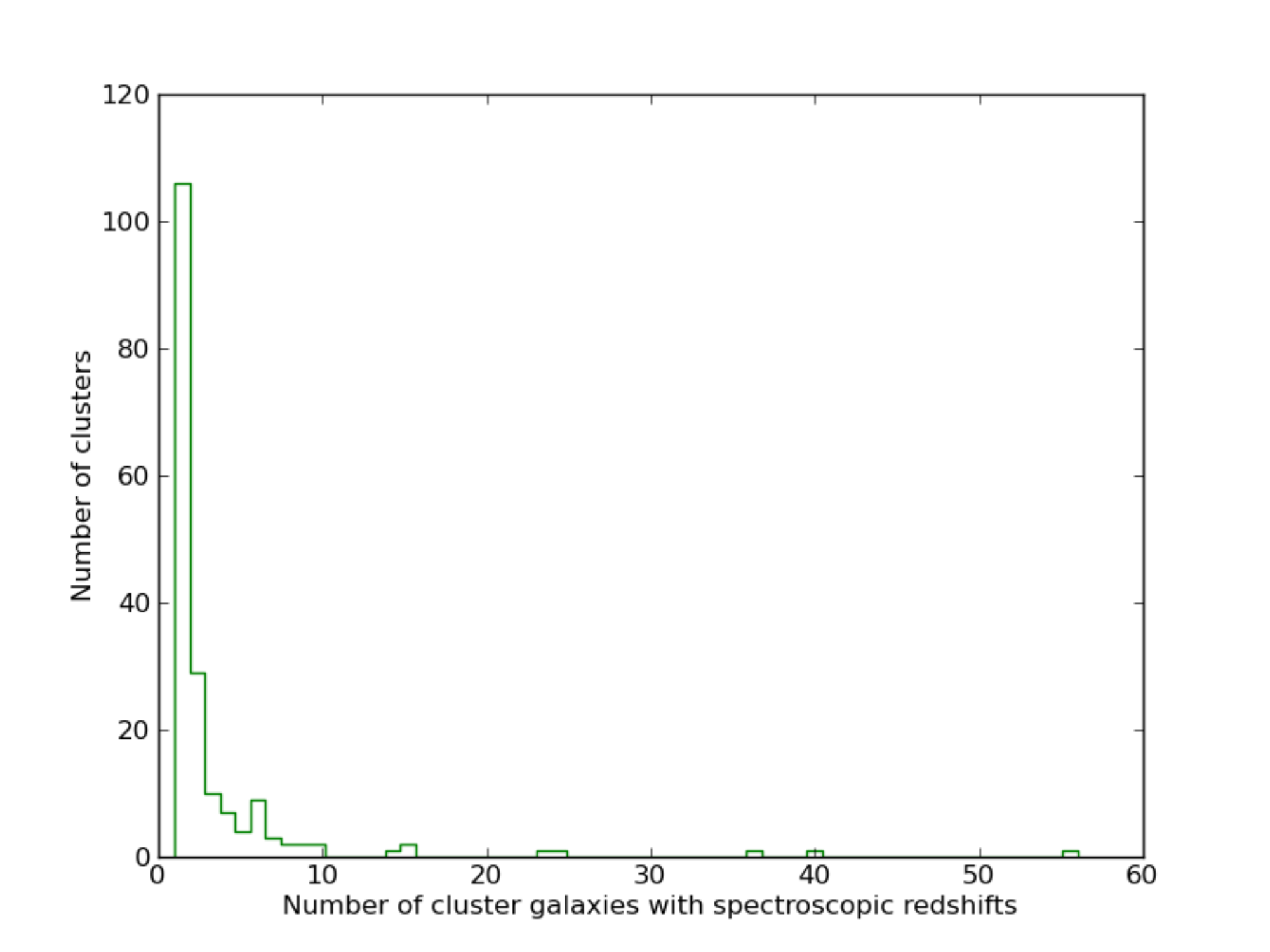}}
  \caption{The distribution of the spectroscopically confirmed cluster members
    per cluster.}
  \label{f:numspec}
\end{figure}

\begin{figure}
  \resizebox{\hsize}{!}{\includegraphics{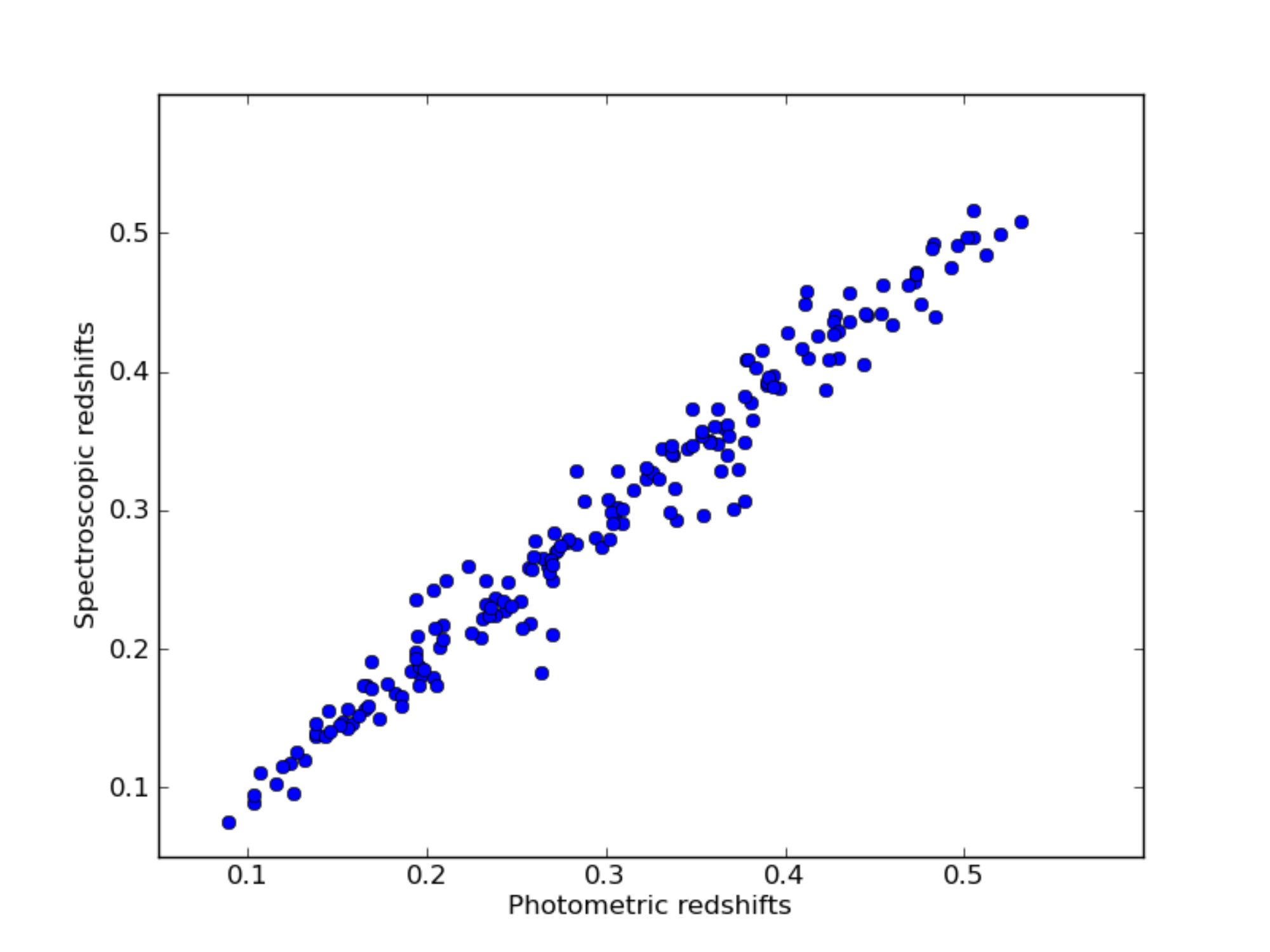}}
  \caption{The relation between the photometric and spectroscopic redshifts of
    the confirmed cluster sample.} 
  \label{f:zpzs}
\end{figure}

The spectroscopic redshift of the cluster was calculated as the average
redshift for the cluster galaxies with spectroscopic redshifts.  The confirmed
cluster sample with spectroscopic redshifts for at least one galaxy 
includes 182 objects. Therefore, the unique optically confirmed X-ray cluster
 sample has the photometric redshifts for all of them, 120 spectroscopic 
 redshifts for 120 BCGs from the optical cluster catalogues, and 182 clusters 
with one or more members with spectroscopic redshifts from the SDSS database. 
Figure~\ref{f:zdist} shows the distribution of the cluster photometric 
redshift $z_{p}$, the distribution of spectroscopic redshifts $z_{s}$ 
of the BCGs as given in the various optical cluster catalogues, and the average
spectroscopic redshift of the cluster members (which we refer to as the cluster
spectroscopic redshift) for the confirmed cluster sample. Figure~\ref{f:numspec} 
shows the distribution of the number of cluster galaxies that have a spectroscopic 
redshift in the SDSS database. The relation between the photometric and spectroscopic 
redshifts of the cluster sample is shown in Figure~\ref{f:zpzs}. Since this relation 
was found to be tight (where the Gaussian distribution of ($z_{s}$ - $z_{p}$) has 
$\sigma$ = 0.02), we were able to rely on the photometric redshifts for the cluster 
with no spectroscopic information.

\begin{figure}
  \resizebox{\hsize}{!}{\includegraphics{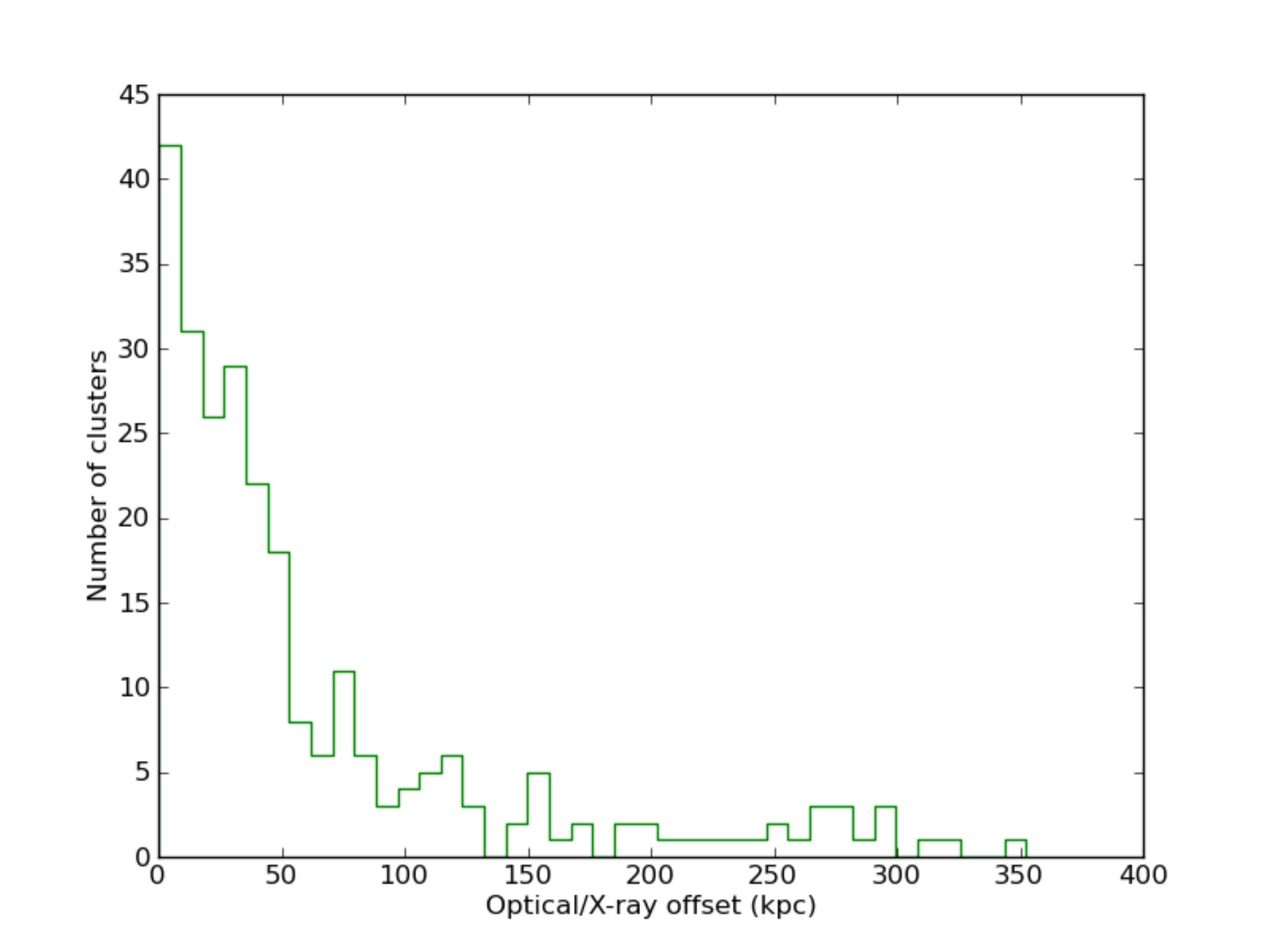}} 
  \caption{ The distribution of the linear separation between the position of
    the BCG and the X-ray cluster position.} 
  \label{f:xoff}
\end{figure}

We used an angular separation of one arcmin to cross-match the X-ray
cluster candidates with the optical cluster catalogues. The corresponding
linear separation was calculated using the spectroscopic redshift, if
available, or the photometric redshift. Figure~\ref{f:xoff} shows the
distribution of the linear separation between the X-ray centre and the BCG
position. For AMF clusters (60 objects), we identified the BCGs of 40 
systems within one arcmin and computed their offsets, which are included in this 
aforementioned distribution. The BCGs were selected as the brightest galaxies with 
$|z_{p} - z_{p(BCG\ cand.)}| < 0.05$ among the three 
BCG candidates given for each AMF cluster published by 
\cite{Szabo11}. The other 20 AMF clusters are not included in 
Fig.~\ref{f:xoff}, because their BCG is outside one arcmin. It is not always the
case that the BCG lies exactly on the X-ray peak. \citet{Rykoff08} model the
optical/X-ray offset distribution by matching a sample of maxBCG clusters to
known X-ray sources from the ROSAT survey. They found a large excess of X-ray
clusters associated with the optical cluster centre. There is a tight core in
which the BCG is within $\sim$ 150\ h$^{-1}$\ kpc  of any X-ray source, as
well as a long tail extending to  $\sim$ 1500\ h$^{-1}$\ kpc. It is shown in 
Figure~\ref{f:xoff} that the majority of the confirmed sample have the BCG 
within a radius ($\sim$ 150 kpc), as well as a tail extending to 352 kpc 
that is consistent with the optical/X-ray offset distribution of   
\cite{Rykoff08}.  

We searched the Astronomical Database SIMBAD and the NASA/IPAC
Extragalactic Database (NED) to check whether they had been identified and catalogued
previously. We used a search radius of one arcmin. About 85 percent of the
confirmed sample are new X-ray clusters, while the remainder had been
previously studied using ROSAT, Chandra, or XMM-Newton data.


\section{X-ray data analysis}
\begin{figure}
  \resizebox{\hsize}{!}{\includegraphics{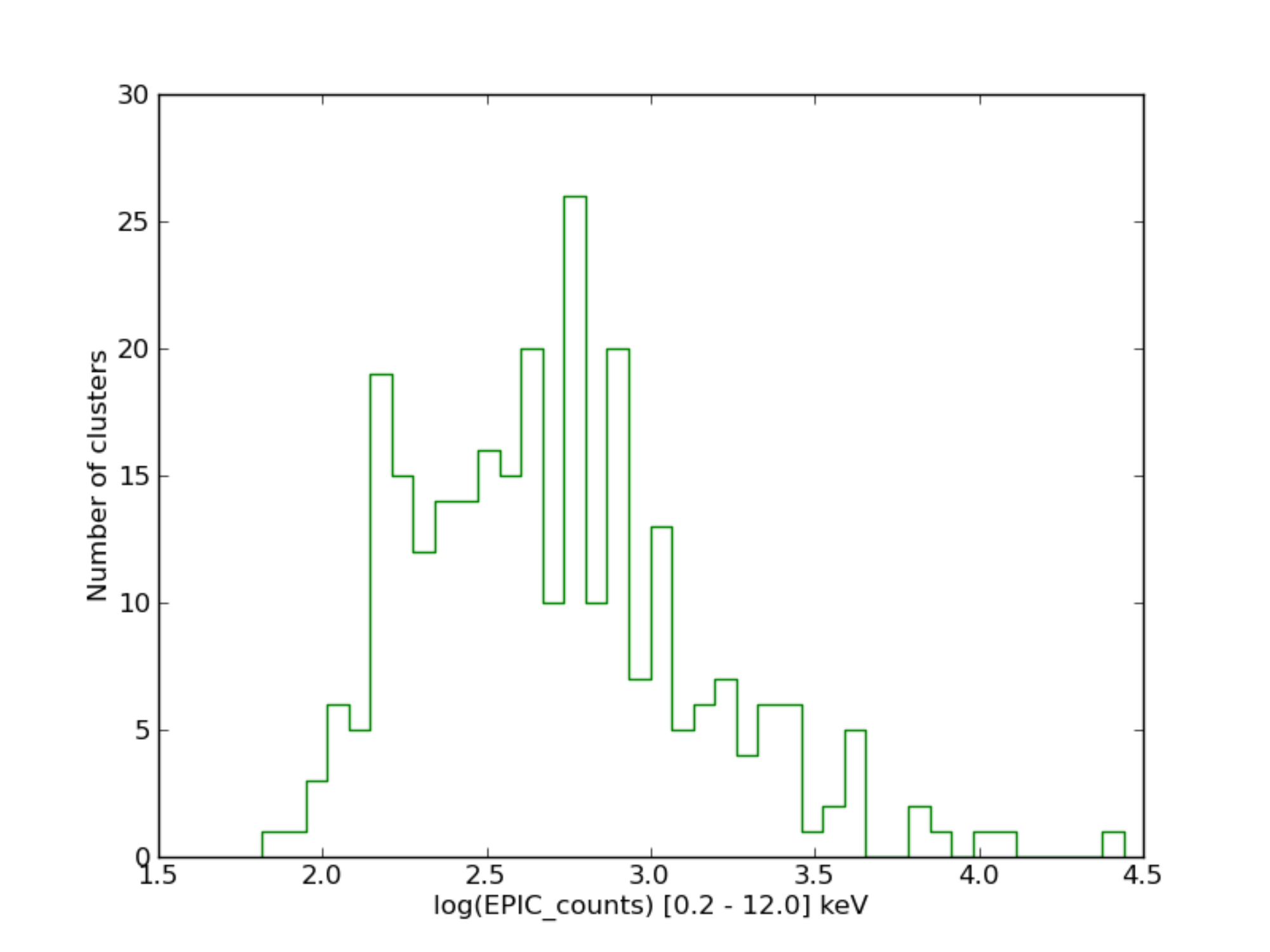}}
  \caption{The distribution of EPIC counts in [0.2-12] keV as given in the 2XMMi-DR3 
catalogue for the confirmed cluster sample.} 
  \label{f:ctsdist}
\end{figure}

The optically confirmed clusters have a wide range of source counts (EPIC counts in
the broad band energy 0.2-12 keV from the 2XMMi-DR3 catalogue) from 66 to 28000 counts 
as shown in Figure~\ref{f:ctsdist}. To analyse
the X-ray data, we have to determine the optical redshifts, except for some   
candidates with more than 1000 net photons for which it is possible to estimate the 
X-ray redshift \citep[e.g.][]{Lamer08,Yu11}. In this paper, we use the cluster 
spectroscopic redshifts where available or the photometric redshifts 
that we obtained from the cross-matching as described in the previous section.

The data reduction and analysis of the optically confirmed sample was carried
out using the XMM-Newton Science Analysis Software (SAS) version 10.0.0. 

\subsection{Standard pipelines}
The raw XMM-Newton data were downloaded using the Archive InterOperability
System (AIO), which provides access to the XMM-Newton Science Archive (XSA). The raw
data were provided in the form of a bundle of files known as observation
data files (ODF), which contain uncalibrated event files, satellite attitude
files, and calibration information. The main steps in the data reduction were:
(i) the generation of calibrated event lists for the EPIC (MOS1, MOS2, and PN)
cameras using the latest calibration data. This was done using the SAS packages
{\tt cifbuild},  {\tt odfingest}, {\tt epchain}, and {\tt emchain}. (ii) The
creation of background light curves to identify time intervals with poor 
quality data. (iii) The filtering of the EPIC event lists to exclude periods of
high background flaring and bad events. (iv) To create a sky image of the
filtered data set. The last three steps were performed using SAS packages {\tt evselect},
{\tt tabgtigen}, and {\tt xmmselect}. 


\subsection{Analysis of the sample}
We now describe the procedure to determine the source and
background regions for each cluster, extract the  source and background
spectra, fit the X-ray spectra, and finally measure the X-ray parameters
(e.g.~temperature, flux, and luminosity). As input to the task generating the X-ray
spectra, we used the filtered event lists as described in the previous
section. 
 

\subsubsection{Optimum source extraction radius}
The most critical step in generating the cluster X-ray spectra is to determine
the source extraction radius. We developed a method to  optimize the
signal-to-noise ratio (SNR) of the spectrum for each cluster.  To
calculate the extraction radius with the highest integrated SNR we created
radial profiles of each cluster in the energy band $0.5 -
2.0$\,keV. Background sources, taken from the EPIC PPS source lists, were
excluded and the profiles were exposure corrected using the EPIC exposure
maps. Since we did not perform a new source detection run, the SNR was 
calculated as a function of radius taking into account the background 
levels as given in the 2XMMi catalogue.

\begin{figure}
  \resizebox{\hsize}{!}{\includegraphics[viewport=0  0 540 392,clip]{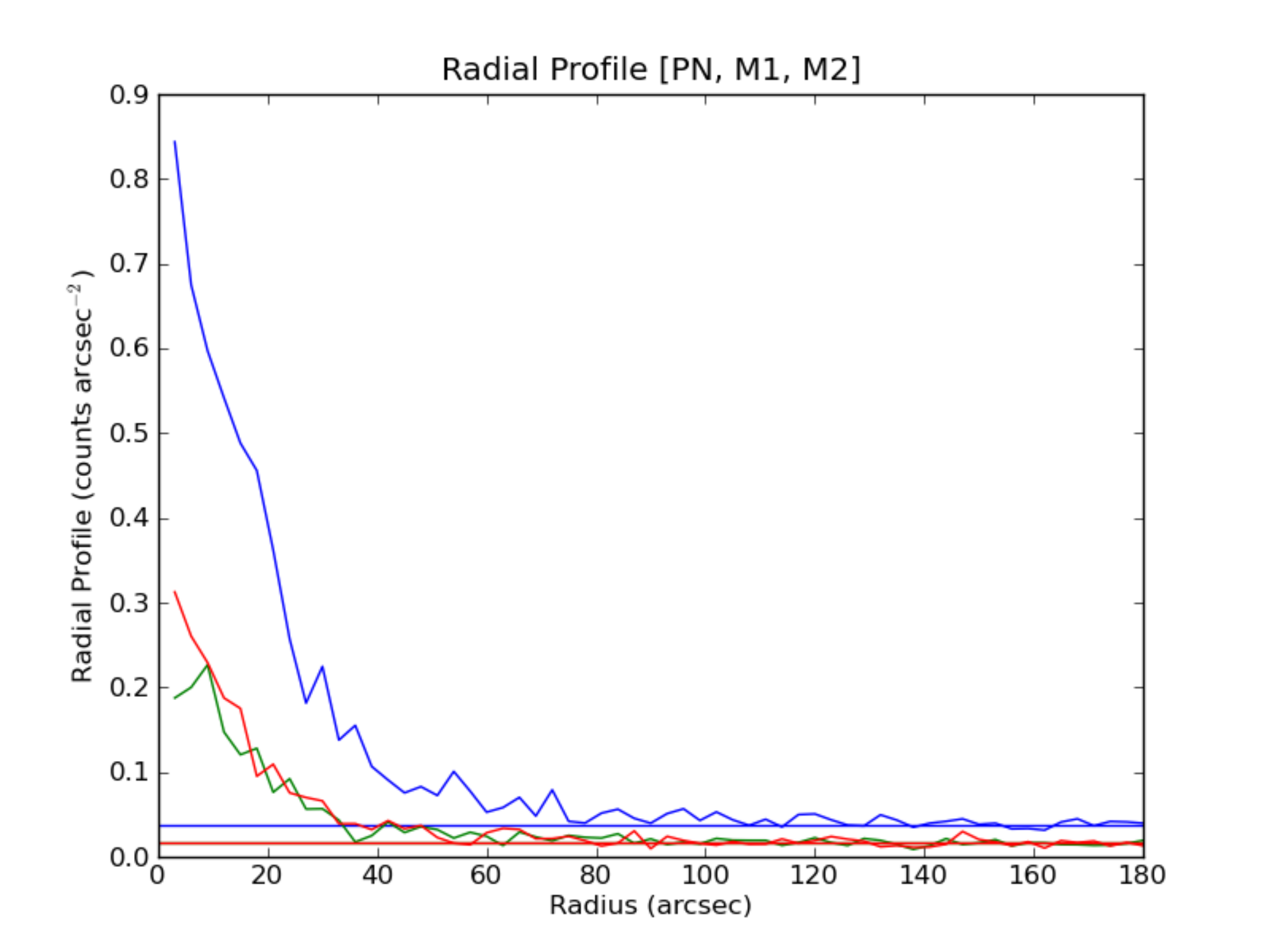}}
  \caption{The radial profile of 2XMM J104421.8+213029 MOS1(green), MOS2(red),
    and PN (blue) images. The horizontal lines indicate the background values
    for MOS1, MOS2 and PN with the same colour as the profile. } 
  \label{f:rprof}
\end{figure}

\begin{figure}
  \resizebox{\hsize}{!}{\includegraphics[viewport=0  0 370 270,clip]{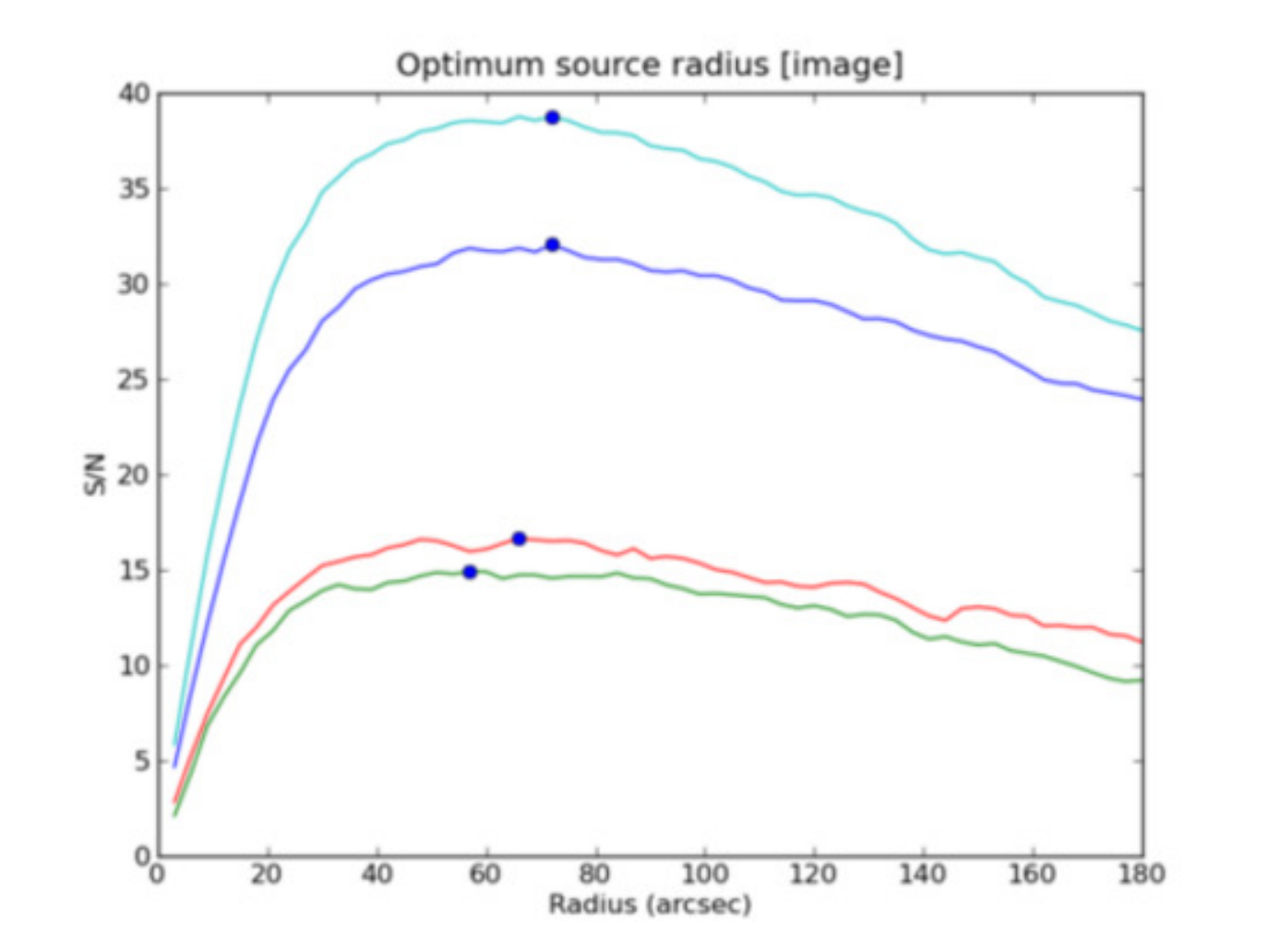}}
  \caption{The signal-to-noise ratio (SNR) profiles of 2XMM J104421.8+213029 in
    MOS1 (green), MOS2 (red), PN (blue) and EPIC (MOS1+MOS2+PN) (cyan) data.
    The cluster optimum extraction radius ($72''$) is corresponding to the
    highest SNR as indicated by a point in the EPIC SNR profile.} 
  \label{f:snr}
\end{figure}

The radial profiles of the X-ray surface brightness of the representative
cluster in MOS1, MOS2 and PN data are shown in Fig.~\ref{f:rprof}. The
background values of the cluster in the EPIC images are indicated by the
horizontal line with the same colours as the profiles. Figure~\ref{f:snr} shows
the SNR profiles of the representative cluster in MOS1, MOS2, PN and EPIC
(MOS1+MOS2+PN) data as a function of the radius from the cluster centre. The
optimum extraction radius ($72''$) is determined from the maximum value in the
EPIC SNR plot, which is indicated by a point in Fig.~\ref{f:snr}. 

\subsubsection{Spectral extraction} 
\begin{figure}
  \resizebox{\hsize}{!}{\includegraphics{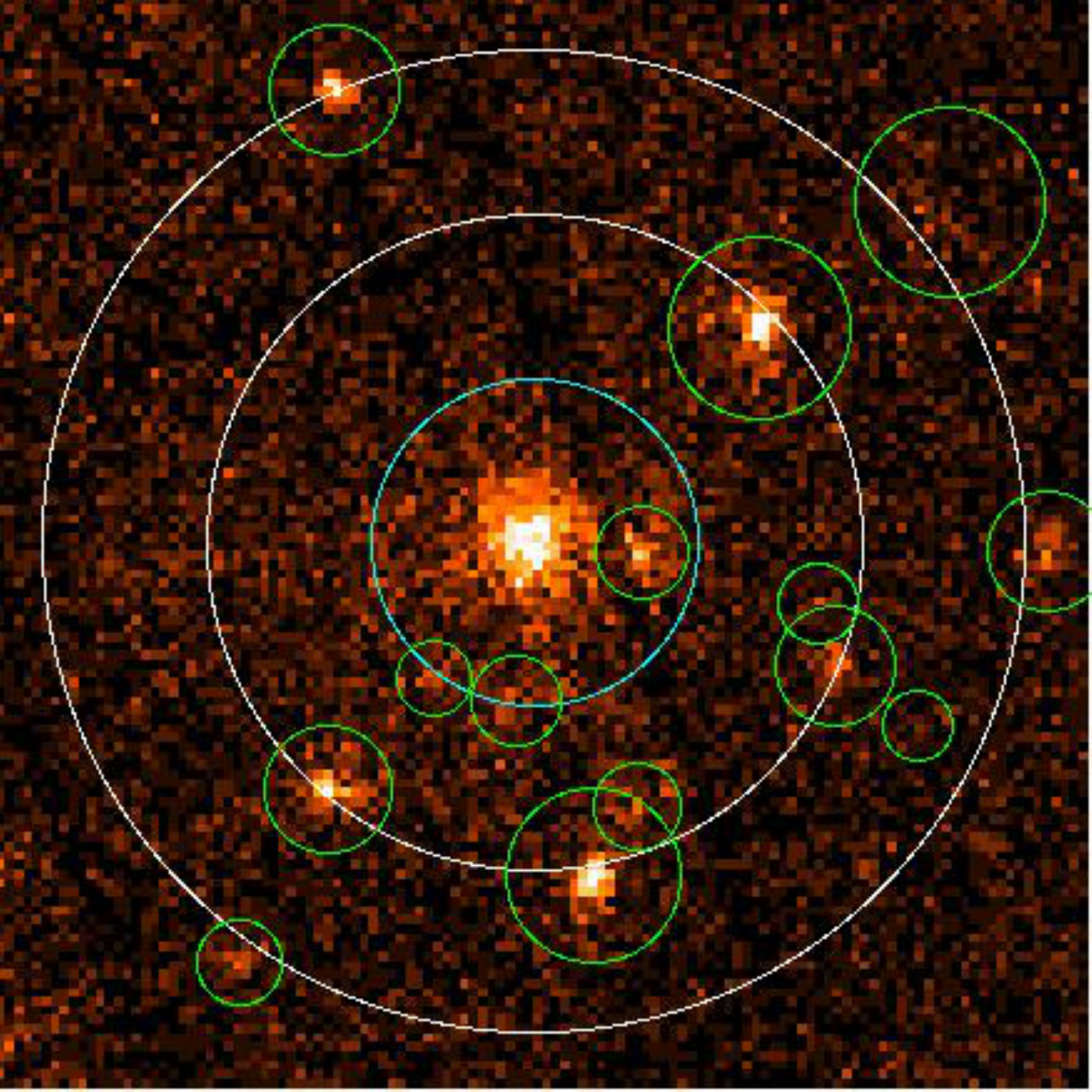}}
  \caption{The representative cluster extraction region is the inner circle
    with colour cyan. The background region is the annulus with white colour.
    The excluding field sources are indicated by green circles. The field of
    view is $8'\times 8'$ centred at the cluster position. } 
  \label{f:mask}
\end{figure}

The EPIC filtered event lists were used to extract the X-ray spectra of the
cross-correlated X-ray optical cluster sample. The spectra of each cluster candidate 
were extracted from a region with an optimum extraction radius as described
in the previous section. The background spectra were extracted from a
circular annulus around the cluster with inner and outer radii equalling two
and three times the optimum radius, respectively. Other unrelated nearby
sources were masked and excluded from the source and background regions that
were finally used to extract the X-ray spectra. 
Figure~\ref{f:mask} shows the cluster and background regions, as well as
the excluded regions of field sources for the representative cluster. The SAS
task {\tt especget} was used to generate the cluster and background spectra and
to create the response matrix files (redistribution  matrix file (RMF) and
ancillary response file (ARF)) required to perform the X-ray spectral
fitting with XSPEC.

\subsubsection{Spectral fitting}
The photon counts of each cluster spectrum were grouped into bins with at least
one count per bin before a fit of a spectral model was applied to the data
using the Ftools task {\tt grppha}. The spectral fitting was carried out using
XSPEC software version 12.5.1 \citep{Arnaud96}. Before executing the algorithm to fit
the spectra, the Galactic HI column (nH) was derived from the HI map from the
Leiden/Argentine/Bonn (LAB) survey \citep{Kalberla05}. This parameter  was fixed
while fitting the X-ray spectrum. The redshift of the spectral
model was fixed to the optical cluster redshift either the spectroscopic redshift
for 182 clusters or the photometric redshift for the remainder cluster sample.  

For each cluster, the available EPIC spectra were fitted simultaneously. The
employed fitting model was a multiplication of a $\tt TBABS$ absorption model 
\citep{Wilms00} and a single-temperature optically thin thermal plasma
component \citep[the $\tt MEKAL$ code   in XSPEC terminology,][]{Mewe86} 
to model the X-ray plasma emission from the ICM. 
The metallicity was fixed at 0.4 $Z\odot$.
 This value is the mean of the metallicities of  95 galaxy clusters
in the redshift range from 0.1 to 0.6 (the same redshift range of the confirmed
sample) observed by Chandra \citep{Maughan08}. The free parameters are the
X-ray temperature and the spectral normalization. The fitting was done using
the Cash statistic with one count per bin  following the recommendation of
\cite{Krumpe08} for small count statistics.  

\begin{figure}
  \resizebox{\hsize}{!}{\includegraphics[angle=-90, viewport=30  0  520 660
      ,clip]{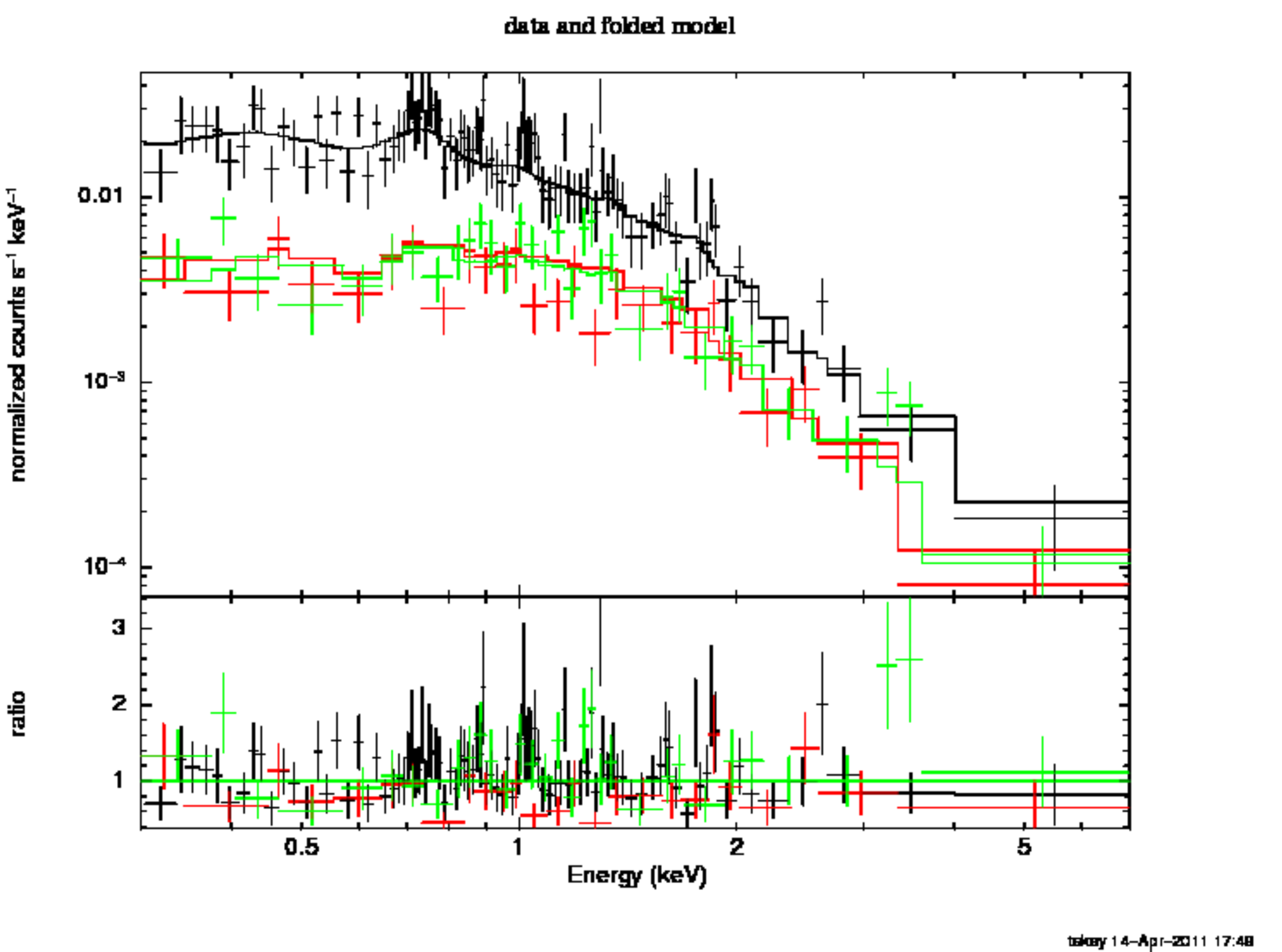}} 
  \caption{The EPIC PN (black), MOS1 (green) and MOS2 (red) data with the best-fit
 MEKAL model for the representative cluster. }  
  \label{f:xfit}
\end{figure}

To avoid the fitting algorithm converging to a local minimum
of the fitting statistics, we ran series of fits stepping from 0.1
to 15 keV  with a step size = 0.05 using the $\tt{steppar}$ command within
XSPEC. The cluster temperature, its flux in the [0.5-2]\,keV band, its X-ray 
luminosity in the [0.5-2]\,keV band, the bolometric luminosity, and the corresponding 
errors were derived from the best-fitting model. We assumed that the fractional error
in the bolometric luminosity was the same as the fractional error in the aperture
luminosity [0.5-2] keV (within an aperture defined by the optimum extraction radius). 
Figure~\ref{f:xfit} shows the fits to the EPIC (MOS1, MOS2, and PN)
spectra and the models for the representative cluster. Figures A.1 to A.4
 in the Appendix A show the fitted spectra of four clusters with different X-ray 
surface brightnesses  and data qualities at different redshifts covering the whole 
redshift range of the confirmed clusters.

\section{Analysis of a cluster sample with reliable X-ray parameters}

We analysed the X-ray data of the optically and X-ray 
confirmed clusters to measure the average temperature of the hot ICM. 
We developed an optimal extraction method for the X-ray spectra
maximising the SNR. The cluster spectra 
were fitted with absorbed thin thermal plasma emission models with
pre-determined redshift and interstellar column density to determine 
the aperture X-ray temperature ($T_{ap}$), flux ($F_{ap}$) [0.5-2] keV, 
luminosity ($L_{ap}$) [0.5-2] keV, and their errors.  
We accepted the measurements of $T_{ap}$ and  $L_{ap}$ if the fractional
errors were smaller than 0.5. About 80 percent of the confirmed clusters passed
this fractional error filter. For these clusters, another visual screening of the
spectral fits (Figure~\ref{f:xfit}) and the X-ray images (Figure~\ref{f:mask}) 
was done.  When the spectral extraction of a given cluster was strongly affected
by the exclusion of field sources within the  extraction radius or a poor 
determination of the background spectrum, it was also excluded from the final 
sample, which comprises 175 clusters. For a fraction of 80 percent, this is 
the first X-ray detection and the first temperature measurement. 

\begin{figure}
  \resizebox{\hsize}{!}{\includegraphics{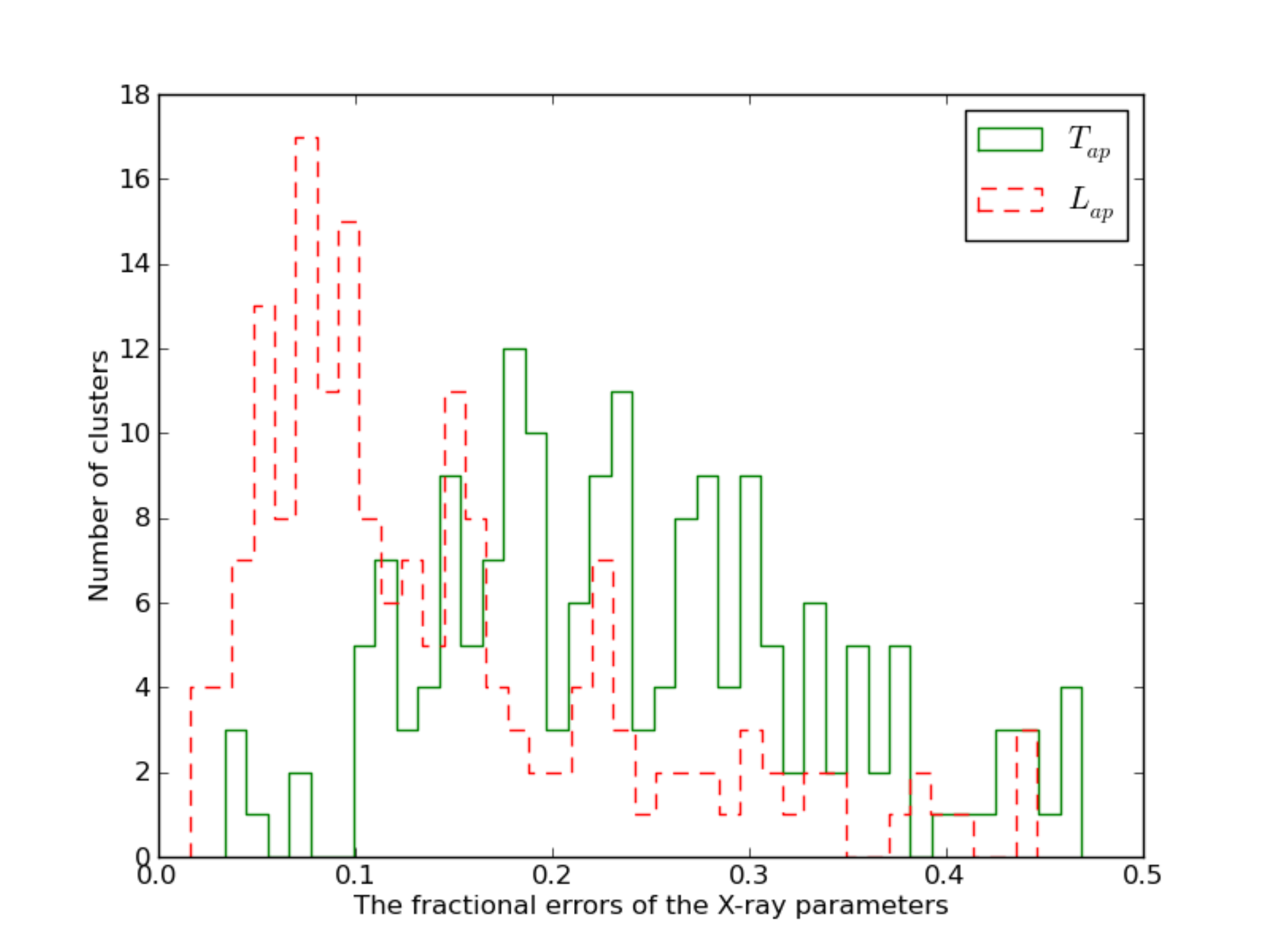}}
  \caption{The distribution of the fractional errors in the X-ray temperatures
    (solid) and luminosities (dashed) derived from spectra extracted within the  
optimum aperture for the energy range 0.5-2 keV of the first cluster sample.}
  \label{f:reler}
\end{figure}

Our subsequent presentation of our analysis and discussion refers to those 175 
objects with reliable X-ray
parameters. The distribution of the $T_{ap}$ and $L_{ap}$ [0.5-2] keV fractional
errors for the first cluster sample is shown in Fig.~\ref{f:reler}. It is
clearly evident that the  cluster luminosity is more tightly constrained than the temperature. 
For about 86 percent of the sample,  the fractional errors 
are smaller than 0.25. Therefore, we estimated  several physical parameters
for each cluster based on the bolometric luminosity $L_{bol}$ within the
optimal aperture. The median correction factor between aperture bolometric  luminosities 
and aperture luminosities in the energy band [0.5-2] keV ( $L_{bol}$ / $L_{ap}$ ) 
was found to be 1.7. We assumed that the fractional error in $L_{bol}$ was identical 
to that of $L_{ap}$ [0.5-2] keV. The estimated parameters are $R_{500}$
the radius at which the mean mass density is 500 times the critical density of
the Universe (see Eq. 2) at the cluster redshift, $L_{500}$ the bolometric 
luminosity within $R_{500}$, and $M_{500}$ the cluster mass within
$R_{500}$. We used an iterative procedure to estimate the physical parameters
using published $L_{500} - T_{500}$ and $L_{500} - M_{500}$ relations  
\citep[][their orthogonal fit for $M_{500}$ with Malmquist bias correction]{Pratt09}.
Our procedure is similar to that used by \cite{Piffaretti10} and \cite{Suhada10},  
which consists of the following steps:

\begin{itemize}
\item[(i)] We estimate $M_{500}$ using the $L - M$ relation
\begin{equation}
 M_{500} = 2 \times 10^{14} M_{\odot}\ \bigl(
 \frac{h(z)^{-7/3}\ L_{bol}}{1.38\times10^{44}\ erg\ s^{-1}} \Bigr)^{1/2.08},  
\end{equation}
where $h(z)$  is the Hubble constant normalised to its present-day value, $
h(z) = \bigl[\Omega_{\rm M} (1+z)^{3} + \Omega_{\Lambda}\Bigr]^{1/2} $. We 
approximate $L_{500}$ as the aperture bolometric luminosity 
$L_{bol}$,  which we correct in an iterative way. 

\item[(ii)] We compute $R_{500}$ 
\begin{equation}
 R_{500} = \sqrt[3]{3 M_{500} / 4\pi 500 \rho_{c}(z)},
\end{equation} 
where the critical density is  $\rho_{c}(z) = h(z)^{2} 3 H^{2} / 8 \pi G$ .

\item[(iii)] 
We compute the cluster temperature within $R_{500}$ using the $L - T$ relation 
\begin{equation}
 T = 5 \mbox{keV} \ \bigl(\frac{h(z)^{-1}\ L_{bol}}{7.13 \times 10^{44}\ erg\ s^{-1}}\Bigr)^{1/3.35}.
\end{equation} 

\item[(iv)] We calculate the core radius $r_{core}$ and $\beta$  using scaling
  relations from \cite{Finoguenov07} 
\begin{equation}
 r_{core}  = 0.07 \times R_{500} \times \bigl(\frac{T}{1\ keV}\bigr)^{0.63},
\end{equation} 
\begin{equation}
 \beta  = 0.4\ \bigl(\frac{T}{1\ keV}\bigr)^{1/3}.
\end{equation} 

\item[(v)] We calculate the enclosed flux within $R_{500}$ and the optimum
  aperture by extrapolating the $\beta$-model. The ratio of the two
  fluxes is calculated, i.e.  $\gamma = F_{500} / F_{\rm bol} $. 

\item[(vi)]
We finally compute a corrected value of $L_{500} = \gamma \times L_{\rm bol}$.
\end{itemize}

We then considered $L_{500}$ as input for another iteration and all
computed parameters were updated. We repeated this iterative procedure until
 converging to a final solution.  At this stage, the $L_{500}$, $M_{500}$, 
and $R_{500}$ were determined. The median correction factor between extrapolated 
luminosities and  aperture bolometric luminosities ($L_{500}$/$L_{\rm bol}$) 
was 1.5. To calculate the errors in Eqs. 1 and 3,  we 
included the measurement errors in the aperture bolometric luminosity $L_{bol}$   
, the intrinsic scatter in  the $L - T$ and $L - M$ relations, and the 
propagated errors caused by the uncertainty in their slopes and intercepts. 
For Eq. 4 and Eq. 5, we included only the propagated errors of their 
independent parameters since their intrinsic scatter had not been published. 
Finally, all the measured errors were taken into account when computing the errors 
in $L_{500}$ and $M_{500}$ in the last iteration. The errors 
in $L_{500}$ and $M_{500}$ were still underestimated because of the possible scatter 
in the relations in Eq. 4 and Eq. 5.       

\begin{figure}
  \resizebox{\hsize}{!}{\includegraphics{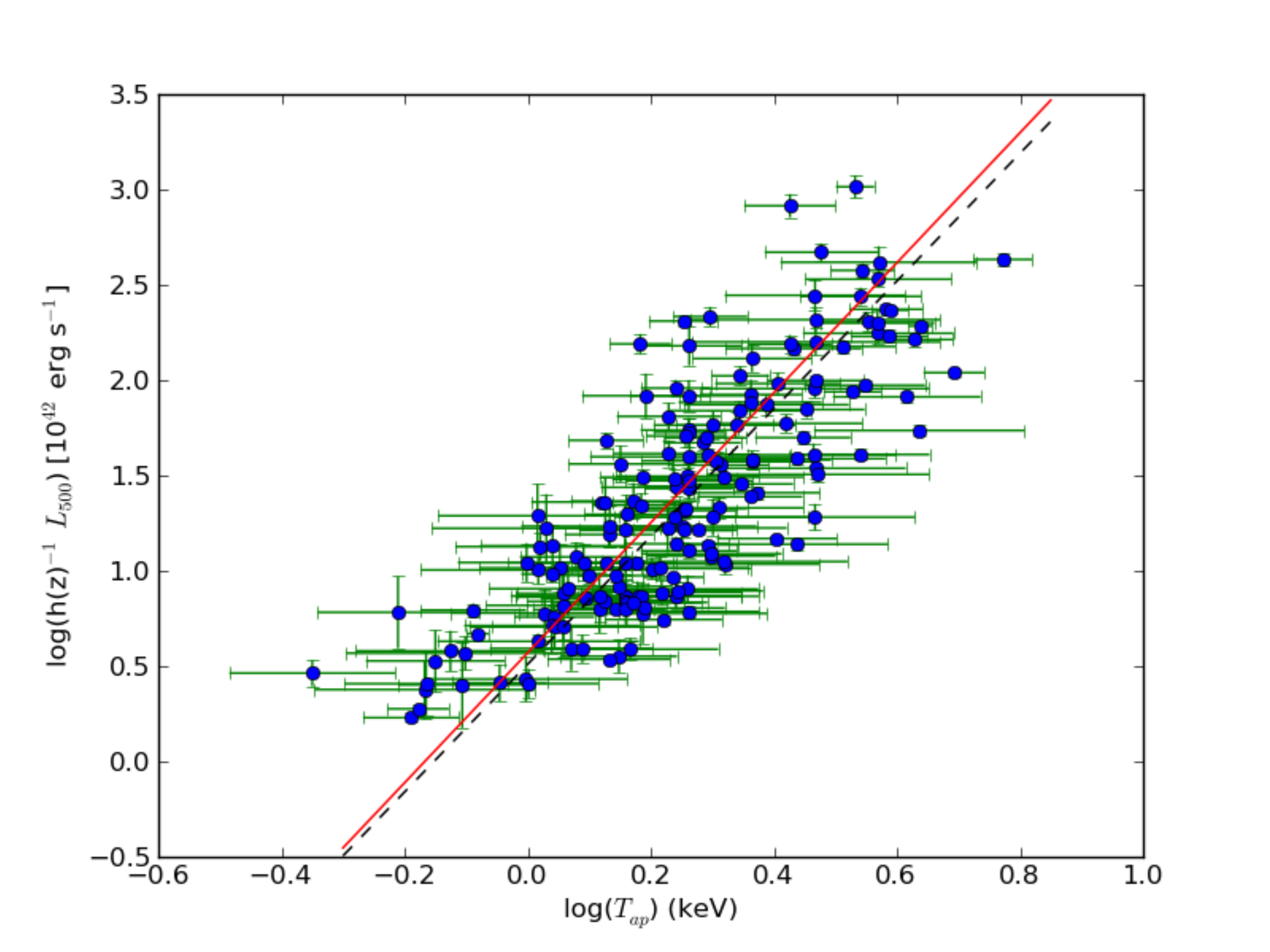}}
  \caption{The relation between the X-ray bolometric luminosities $L_{500}$ and 
aperture temperatures $T_{ap}$ of the first cluster sample.  The solid line indicates
 the best fit of the sample using orthogonal distance regression (ODR). The dashed 
line is the extrapolated relation for REXCESS sample \citep{Pratt09} using 
a BCES orthogonal fit.}
  \label{f:L-T}
\end{figure}

We investigated the $L - T$ relation in the first cluster sample using $T_{ap}$ and
$L_{500}$.  Figure~\ref{f:L-T} shows the relation between $L_{500}$ (corrected for the
redshift evolution) and $T_{ap}$ (uncorrected for cooling flows).  Here we
assumed that $T_{ap}$ did not differ significantly from $T_{500}$ and its
error was derived from the spectral fits. The best-fit linear relation (solid line) 
derived from an orthogonal distance  regression fit (ODR) \citep[][which takes 
into account measurement errors in both variables]{Boggs90} between their logarithm, is 

\begin{equation}
\log\ (h(z)^{-1}\ L_{500}) = (0.57 \pm 0.05)  + (3.41 \pm 0.15)\ \log\ (T_{ap}).
\end{equation} 

The best-fit power law relation derived from a BCES orthogonal fit to the
$L_{500}$ - $T_{500}$ relation published by \cite{Pratt09}  for the REXCESS
sample is plotted as the dashed line in Fig.~\ref{f:L-T}. 
The ODR slope (present work), $3.41 \pm 0.15$, is consistent with the BCES 
orthogonal slope \citep{Pratt09} of the REXCESS
sample, $3.35 \pm 0.32$. In addition, the present slope is consistent with the BCES 
orthogonal slope (3.63 $\pm$ 0.27) of the $L - T$ relation derived from a sample of 
114 clusters (without excluding the core regions) observed 
with Chandra across a wide range of temperature (2 $<$ kT $<$ 16 keV) 
and redshift (0.1 $<$ z $<$ 1.3) by \cite{Maughan11}. 

We tested the corresponding uncertainty in the error budget of $L_{500}$ caused 
by the above-mentioned unknown scatter in Eq. 4 and 5: for example, a $\sigma = \pm 0.1$ 
scatter in the $\beta$ value results in a fractional error in   $L_{500}$ of 17$\%$. 
If we take into account the newly estimated errors in $L_{500}$ when fitting the $L - T$ relation, 
the revised slope of 3.32 is within the error in the original slope as in Eq. 6 and still 
consistent with the published ones.
   
\begin{figure}
  \resizebox{\hsize}{!}{\includegraphics{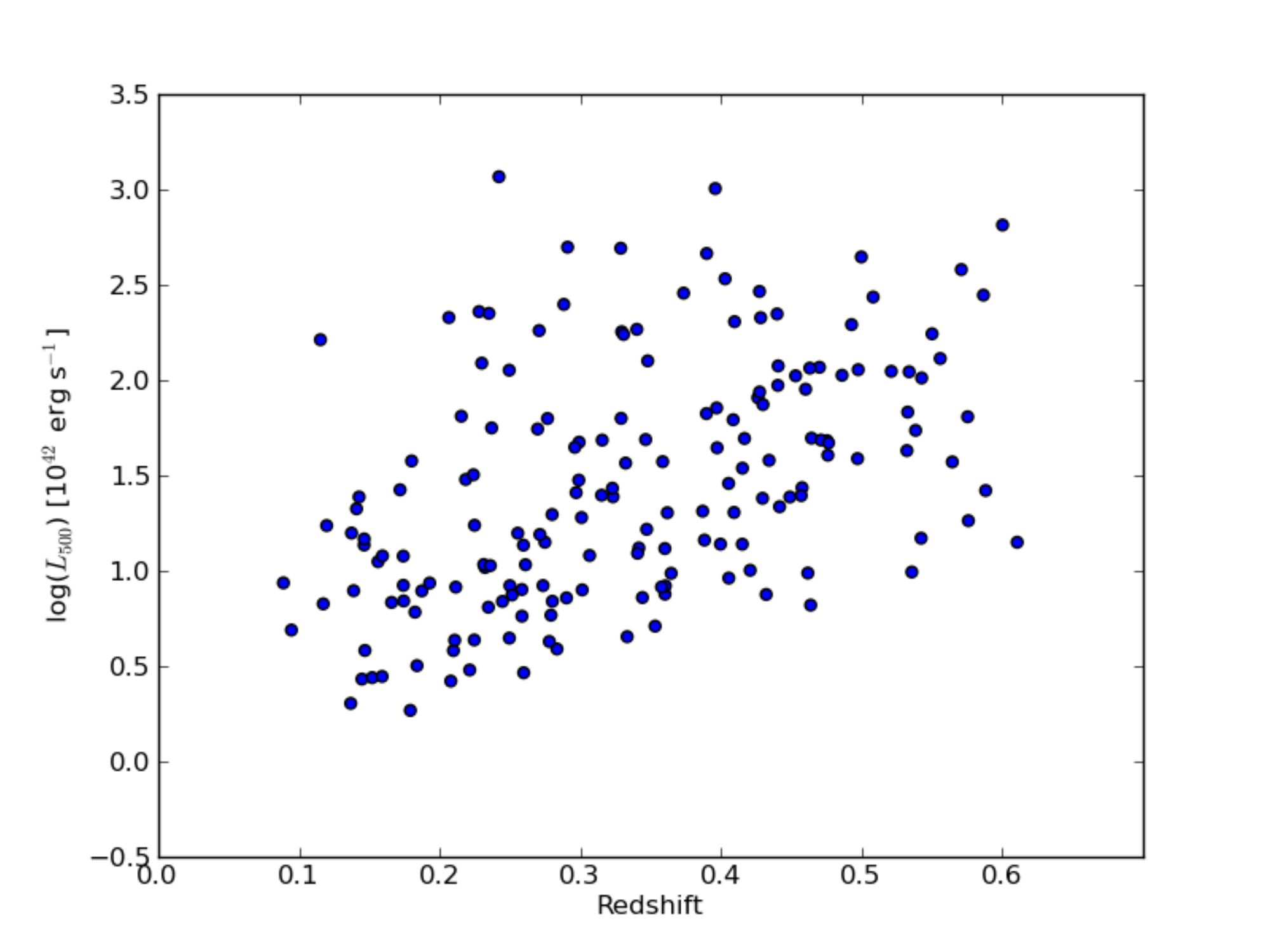}}
  \caption{The distribution of the X-ray bolometric luminosities $L_{500}$ as a 
function of optical redshifts of the first cluster sample.}
  \label{f:L-z}
\end{figure}

Our sample represents cluster temperatures ranging from 0.45 to 5.92 keV and values of 
bolometric luminosity in the $L_{500}$ range $1.9 \times 10^{42} - 1.2 \times
10^{45}$\ erg\ s$^{-1}$  in a wide redshift range  0.1 - 0.6.  Most of the
published $L - T$ relations were derived from   local cluster samples with
temperatures higher than 2 keV. The current relation is derived for our
sample, which includes clusters and  groups with low temperatures 
and luminosities in a wide redshift range up to $z = 0.6$. The distribution of
luminosity as a function of redshift is shown in Fig.~\ref{f:L-z}.

Table 2, available at the CDS,  represents the first cluster sample containing as many as  
175 X-ray clusters. In addition, the first cluster sample with the X-ray-optical overlay and fitted spectra 
for each cluster is publicly available from http://www.aip.de/groups/xray/XMM$_{-}$SDSS$_{-}$CLUSTERS.
 In the catalogue, we provide the cluster identification number 
(detection Id, detid) and its name (IAUNAME) in (cols. [1] and [2]), the right 
ascension and declination of X-ray emission in equinox J2000.0 (cols. [3] and [4]), 
the XMM-Newton observation Id (obsid) (col. [5]), the optical redshift (col. [6]), 
the scale at the cluster redshift in kpc/$''$ (col. [7]), 
the aperture and $R_{500}$ radii in kpc (col.[8] and [9]), the cluster aperture  
X-ray temperature $T_{ap}$ and its positive and negative errors in keV (cols. [10], 
[11] and [12], respectively), the aperture X-ray flux $F_{ap}$ [0.5-2] keV 
and its positive and negative errors in units of $10^{-14}$\ erg\ cm$^{-2}$\ s$^{-1}$ 
(cols. [13], [14] and [15], respectively), the aperture  
X-ray luminosity $L_{ap}$ [0.5-2] keV and its positive and negative errors in units 
of $10^{42}$\ erg\ s$^{-1}$ (cols. [16], [17] and [18], respectively),
the cluster bolometric luminosity $L_{500}$ and its error in units of 
$10^{42}$\ erg\ s$^{-1}$ (cols. [19] and [20]), the cluster mass $M_{500}$ and its error
in units of $10^{13}$\ M$\odot$ (cols. [21] and [22]), the Galactic HI column in 
units $10^{22}$\ cm$^{-2}$ (col.[23]), the identification number of the cluster
in optical catalogue (col.[24]), the BCG right ascension and declination in equinox 
J2000.0 (cols. [25] and [26]) although for AMF catalogue they represent the 
cluster stellar mass centre, the cluster photometric redshift (col.[27]), the average 
spectroscopic redshift of the cluster galaxies with available spectroscopic
redshifts and their number (cols.[28] and [29]), the linear offset between the 
cluster X-ray position and the cluster optical position (col.[30]), 
the optical cluster catalogue names that identify the cluster (col.[31]) 
(Note: the optical parameters are extracted from the first one), 
and the alternative name of the  X-ray clusters previously identified using ROSAT, Chandra, or 
XMM-Newton data and its reference in NED and SIMBAD databases (col.[32] and [33]).


\section{Summary and outlook}
We have presented the first sample of X-ray galaxy clusters from the 2XMMi-Newton/SDSS 
Galaxy Cluster Survey. The survey comprises 1180 cluster candidates selected as X-ray 
serendipitous sources from the second XMM-Newton serendipitous source catalogue
(2XMMi-DR3) that had been observed by the SDSS. A quarter of the candidates 
are identified as distant cluster candidates beyond z = 0.6, because there is no apparent 
overdensity of galaxies in the corresponding 
SDSS images. Another quarter of the candidates had been previously identified
in optical cluster catalogues extracted from SDSS data.    
Our cross-correlation of the X-ray cluster candidates with four optical cluster catalogues within 
a matching radius of one arcmin confirmed 275 clusters and provided us with 
the photometric redshifts for all of them and the spectroscopic redshifts for 120 BCGs. 
We extracted all available spectroscopic redshifts 
for the cluster members from recent SDSS data. Among the confirmed cluster sample, 182 clusters 
have spectroscopic redshifts for at least one galaxy member.   
More than 80 percent of the confirmed sample are newly identified X-ray clusters and the 
others had been previously identified using ROSAT, Chandra, or XMM-Newton data.
 We reduced and analysed the X-ray data of the confirmed sample in an automated way. 
The X-ray temperature, flux and luminosity of
the confirmed sample and their errors were derived from  spectral fitting.
 The analysed sample in the present work contains 175 X-ray galaxy clusters with acceptable
 measurements of X-ray parameters (ie. with fractional errors smaller than 0.5) 
from reasonable quality fitting (139 objects being newly discovered in X-rays).
 In addition, we derived the physical properties ($R_{500}$, $L_{500}$ and $M_{500}$) 
of the study sample from an iterative procedure using the published scaling relations. 
The relation between the X-ray bolometric luminosity $L_{500}$ and aperture temperature of 
the sample is investigated. The slope of the relation agrees with the slope of the 
same relation in the  REXCESS sample \citep{Pratt09}. The present relation is derived from
 a large sample with low luminosities and temperatures  across a wide redshift range 0.09 - 0.61.   
 
As one extension to this project, we intend to obtain SDSS photometric redshifts
 of all 2XMMi-DR3 X-ray cluster candidates that have been detected in  SDSS imaging. This 
will significantly increase the sample size and the identified fraction of the 2XMMi cluster sample.
Further improvements in the accuracy of the X-ray parameters for about  10 percent of the confirmed 
sample will be made by analysing repeated observations of those clusters.


 \begin{acknowledgements}
This work is supported by the Egyptian Ministry of Higher Education and 
Scientific Research in cooperation with the Leibniz-Institut f{\"u}r Astrophysik 
Potsdam (AIP), Germany. We also acknowledge the partial support by the 
Deutsches Zentrum f{\"u}r Luft- und Raumfahrt (DLR) under contract number  
50 QR 0802. We acknowledge the anonymous referee for suggestions that improved 
the discussion of the results. The XMM-Newton project is an ESA Science  Mission 
with instruments and contributions directly funded by ESA Member States and the 
USA (NASA). Funding for SDSS-III has been provided by the Alfred P. Sloan Foundation, 
the Participating Institutions, the National Science Foundation,
 and the U.S. Department of Energy. The SDSS-III web site is http://www.sdss3.org/.
 SDSS-III is managed by the Astrophysical Research Consortium for the Participating Institutions of 
the SDSS-III Collaboration including the
 University of Arizona, the Brazilian Participation Group, Brookhaven National Laboratory, University of 
Cambridge, University of Florida, the French Participation 
Group, the German Participation Group, the Instituto de Astrofisica de Canarias, the Michigan State/Notre 
Dame/JINA Participation Group, Johns Hopkins University,
 Lawrence Berkeley National Laboratory, Max Planck Institute for Astrophysics, New Mexico State University, 
New York University, Ohio State University,
 Pennsylvania State University, University of Portsmouth, Princeton University, the Spanish Participation Group, 
University of Tokyo,
 University of Utah, Vanderbilt University, University of Virginia, University of Washington, and Yale University.
 \end{acknowledgements}

 \bibliographystyle{aa}
 \bibliography{17498}

\begin{thebibliography}{50}
\expandafter\ifx\csname natexlab\endcsname\relax\def\natexlab#1{#1}\fi

\bibitem[{{Arnaud}(1996)}]{Arnaud96}
{Arnaud}, K.~A. 1996, in Astronomical Society of the Pacific Conference Series,
  Vol. 101, Astronomical Data Analysis Software and Systems V, ed.
  {G.~H.~Jacoby \& J.~Barnes}, 17

\bibitem[{{Arnaud} {et~al.}(2010){Arnaud}, {Pratt}, {Piffaretti},
  {B{\"o}hringer}, {Croston}, \& {Pointecouteau}}]{Arnaud10}
{Arnaud}, M., {Pratt}, G.~W., {Piffaretti}, R., {et~al.} 2010, \aap, 517, A92

\bibitem[{{Arviset} {et~al.}(2002){Arviset}, {Guainazzi}, {Hernandez},
  {Dowson}, {Osuna}, \& {Venet}}]{Arviset02}
{Arviset}, C., {Guainazzi}, M., {Hernandez}, J., {et~al.} 2002, ArXiv
  Astrophysics e-prints

\bibitem[{{Barkhouse} {et~al.}(2006){Barkhouse}, {Green}, {Vikhlinin}, {Kim},
  {Perley}, {Cameron}, {Silverman}, {Mossman}, {Burenin}, {Jannuzi}, {Kim},
  {Smith}, {Smith}, {Tananbaum}, \& {Wilkes}}]{Barkhouse06}
{Barkhouse}, W.~A., {Green}, P.~J., {Vikhlinin}, A., {et~al.} 2006, \apj, 645,
  955

\bibitem[{{Basilakos} {et~al.}(2004){Basilakos}, {Plionis}, {Georgakakis},
  {Georgantopoulos}, {Gaga}, {Kolokotronis}, \& {Stewart}}]{Basilakos04}
{Basilakos}, S., {Plionis}, M., {Georgakakis}, A., {et~al.} 2004, \mnras, 351,
  989

\bibitem[{{Boggs} \& {Rogers}(1990)}]{Boggs90}
{Boggs}, P.~T. \& {Rogers}, J.~E. 1990, Contemporary Mathematics, 112, 183

\bibitem[{{B{\"o}hringer} {et~al.}(2002){B{\"o}hringer}, {Collins}, {Guzzo},
  {Schuecker}, {Voges}, {Neumann}, {Schindler}, {Chincarini}, {De Grandi},
  {Cruddace}, {Edge}, {Reiprich}, \& {Shaver}}]{Boehringer02}
{B{\"o}hringer}, H., {Collins}, C.~A., {Guzzo}, L., {et~al.} 2002, \apj, 566,
  93

\bibitem[{{Burenin} {et~al.}(2007){Burenin}, {Vikhlinin}, {Hornstrup},
  {Ebeling}, {Quintana}, \& {Mescheryakov}}]{Burenin07}
{Burenin}, R.~A., {Vikhlinin}, A., {Hornstrup}, A., {et~al.} 2007, \apjs, 172,
  561

\bibitem[{{Cavagnolo} {et~al.}(2008){Cavagnolo}, {Donahue}, {Voit}, \&
  {Sun}}]{Cavagnolo08}
{Cavagnolo}, K.~W., {Donahue}, M., {Voit}, G.~M., \& {Sun}, M. 2008, \apj, 682,
  821

\bibitem[{{Dietrich} {et~al.}(2007){Dietrich}, {Erben}, {Lamer}, {Schneider},
  {Schwope}, {Hartlap}, \& {Maturi}}]{Dietrich07}
{Dietrich}, J.~P., {Erben}, T., {Lamer}, G., {et~al.} 2007, \aap, 470, 821

\bibitem[{{Fassbender} {et~al.}(2007){Fassbender}, {B{\"o}hringer}, {Santos},
  {Schuecker}, {Lamer}, {Schwope}, {Kohnert}, {Rosati}, {Mullis}, \&
  {Quintana}}]{Fassbender07}
{Fassbender}, R., {B{\"o}hringer}, H., {Santos}, J., {et~al.} 2007, in Heating
  versus Cooling in Galaxies and Clusters of Galaxies, ed. {H.~B{\"o}hringer,
  G.~W.~Pratt, A.~Finoguenov, \& P.~Schuecker }, 54

\bibitem[{{Finoguenov} {et~al.}(2007){Finoguenov}, {Guzzo}, {Hasinger},
  {Scoville}, {Aussel}, {B{\"o}hringer}, {Brusa}, {Capak}, {Cappelluti},
  {Comastri}, {Giodini}, {Griffiths}, {Impey}, {Koekemoer}, {Kneib},
  {Leauthaud}, {Le F{\`e}vre}, {Lilly}, {Mainieri}, {Massey}, {McCracken},
  {Mobasher}, {Murayama}, {Peacock}, {Sakelliou}, {Schinnerer}, {Silverman},
  {Smol{\v c}i{\'c}}, {Taniguchi}, {Tasca}, {Taylor}, {Trump}, \&
  {Zamorani}}]{Finoguenov07}
{Finoguenov}, A., {Guzzo}, L., {Hasinger}, G., {et~al.} 2007, \apjs, 172, 182

\bibitem[{{Finoguenov} {et~al.}(2010){Finoguenov}, {Watson}, {Tanaka},
  {Simpson}, {Cirasuolo}, {Dunlop}, {Peacock}, {Farrah}, {Akiyama}, {Ueda},
  {Smol{\v c}i{\'c}}, {Stewart}, {Rawlings}, {van Breukelen}, {Almaini},
  {Clewley}, {Bonfield}, {Jarvis}, {Barr}, {Foucaud}, {McLure}, {Sekiguchi}, \&
  {Egami}}]{Finoguenov10}
{Finoguenov}, A., {Watson}, M.~G., {Tanaka}, M., {et~al.} 2010, \mnras, 403,
  2063

\bibitem[{{Hao} {et~al.}(2010){Hao}, {McKay}, {Koester}, {Rykoff}, {Rozo},
  {Annis}, {Wechsler}, {Evrard}, {Siegel}, {Becker}, {Busha}, {Gerdes},
  {Johnston}, \& {Sheldon}}]{Hao10}
{Hao}, J., {McKay}, T.~A., {Koester}, B.~P., {et~al.} 2010, \apjs, 191, 254

\bibitem[{{Horner} {et~al.}(2008){Horner}, {Perlman}, {Ebeling}, {Jones},
  {Scharf}, {Wegner}, {Malkan}, \& {Maughan}}]{Horner08}
{Horner}, D.~J., {Perlman}, E.~S., {Ebeling}, H., {et~al.} 2008, \apjs, 176,
  374

\bibitem[{{Kalberla} {et~al.}(2005){Kalberla}, {Burton}, {Hartmann}, {Arnal},
  {Bajaja}, {Morras}, \& {P{\"o}ppel}}]{Kalberla05}
{Kalberla}, P.~M.~W., {Burton}, W.~B., {Hartmann}, D., {et~al.} 2005, \aap,
  440, 775

\bibitem[{{Koester} {et~al.}(2007){Koester}, {McKay}, {Annis}, {Wechsler},
  {Evrard}, {Bleem}, {Becker}, {Johnston}, {Sheldon}, {Nichol}, {Miller},
  {Scranton}, {Bahcall}, {Barentine}, {Brewington}, {Brinkmann}, {Harvanek},
  {Kleinman}, {Krzesinski}, {Long}, {Nitta}, {Schneider}, {Sneddin}, {Voges},
  \& {York}}]{Koester07}
{Koester}, B.~P., {McKay}, T.~A., {Annis}, J., {et~al.} 2007, \apj, 660, 239

\bibitem[{{Krumpe} {et~al.}(2008){Krumpe}, {Lamer}, {Corral}, {Schwope},
  {Carrera}, {Barcons}, {Page}, {Mateos}, {Tedds}, \& {Watson}}]{Krumpe08}
{Krumpe}, M., {Lamer}, G., {Corral}, A., {et~al.} 2008, \aap, 483, 415

\bibitem[{{Lamer} {et~al.}(2008){Lamer}, {Hoeft}, {Kohnert}, {Schwope}, \&
  {Storm}}]{Lamer08}
{Lamer}, G., {Hoeft}, M., {Kohnert}, J., {Schwope}, A., \& {Storm}, J. 2008,
  \aap, 487, L33

\bibitem[{{Marriage} {et~al.}(2010){Marriage}, {Acquaviva}, {Ade}, {Aguirre},
  {Amiri}, {Appel}, {Barrientos}, {Battistelli}, {Bond}, {Brown}, {Burger},
  {Chervenak}, {Das}, {Devlin}, {Dicker}, {Doriese}, {Dunkley}, {Dunner},
  {Essinger-Hileman}, {Fisher}, {Fowler}, {Hajian}, {Halpern}, {Hasselfield},
  {Hern'andez-Monteagudo}, {Hilton}, {Hilton}, {Hincks}, {Hlozek},
  {Huffenberger}, {Hughes}, {Hughes}, {Infante}, {Irwin}, {Juin}, {Kaul},
  {Klein}, {Kosowsky}, {Lau}, {Limon}, {Lin}, {Lupton}, {Marsden}, {Martocci},
  {Mauskopf}, {Menanteau}, {Moodley}, {Moseley}, {Netterfield}, {Niemack},
  {Nolta}, {Page}, {Parker}, {Partridge}, {Quintana}, {Reese}, {Reid},
  {Sehgal}, {Sherwin}, {Sievers}, {Spergel}, {Staggs}, {Swetz}, {Switzer},
  {Thornton}, {Trac}, {Tucker}, {Warne}, {Wilson}, {Wollack}, \&
  {Zhao}}]{Marriage10}
{Marriage}, T.~A., {Acquaviva}, V., {Ade}, P.~A.~R., {et~al.} 2010, ArXiv
  e-prints

\bibitem[{{Maughan} {et~al.}(2011){Maughan}, {Giles}, {Randall}, {Jones}, \&
  {Forman}}]{Maughan11}
{Maughan}, B.~J., {Giles}, P.~A., {Randall}, S.~W., {Jones}, C., \& {Forman},
  W.~R. 2011, ArXiv e-prints

\bibitem[{{Maughan} {et~al.}(2008){Maughan}, {Jones}, {Forman}, \& {Van
  Speybroeck}}]{Maughan08}
{Maughan}, B.~J., {Jones}, C., {Forman}, W., \& {Van Speybroeck}, L. 2008,
  \apjs, 174, 117

\bibitem[{{Mewe} {et~al.}(1986){Mewe}, {Lemen}, \& {van den Oord}}]{Mewe86}
{Mewe}, R., {Lemen}, J.~R., \& {van den Oord}, G.~H.~J. 1986, \aaps, 65, 511

\bibitem[{{Pierre} {et~al.}(2006){Pierre}, {Pacaud}, \& {Xmm-Lss
  Consortium}}]{Pierre06}
{Pierre}, M., {Pacaud}, F., \& {Xmm-Lss Consortium}, T. 2006, in ESA Special
  Publication, Vol. 604, The X-ray Universe 2005, ed. {A.~Wilson}, 819

\bibitem[{{Piffaretti} {et~al.}(2010){Piffaretti}, {Arnaud}, {Pratt},
  {Pointecouteau}, \& {Melin}}]{Piffaretti10}
{Piffaretti}, R., {Arnaud}, M., {Pratt}, G.~W., {Pointecouteau}, E., \&
  {Melin}, J. 2010, ArXiv e-prints

\bibitem[{{Planck collaboration} {et~al.}(2011){Planck collaboration}, {Ade},
  {Aghanim}, {Arnaud}, {Ashdown}, {Aumont}, {Baccigalupi}, {Balbi}, {Banday},
  {Barreiro}, \& et~al.}]{Planck11}
{Planck collaboration}, {Ade}, P.~A.~R., {Aghanim}, N., {et~al.} 2011, ArXiv
  e-prints

\bibitem[{{Plionis} {et~al.}(2005){Plionis}, {Basilakos}, {Georgantopoulos}, \&
  {Georgakakis}}]{Plionis05}
{Plionis}, M., {Basilakos}, S., {Georgantopoulos}, I., \& {Georgakakis}, A.
  2005, \apjl, 622, L17

\bibitem[{{Pratt} {et~al.}(2010){Pratt}, {Arnaud}, {Piffaretti},
  {B{\"o}hringer}, {Ponman}, {Croston}, {Voit}, {Borgani}, \&
  {Bower}}]{Pratt10}
{Pratt}, G.~W., {Arnaud}, M., {Piffaretti}, R., {et~al.} 2010, \aap, 511, A85

\bibitem[{{Pratt} {et~al.}(2009){Pratt}, {Croston}, {Arnaud}, \&
  {B{\"o}hringer}}]{Pratt09}
{Pratt}, G.~W., {Croston}, J.~H., {Arnaud}, M., \& {B{\"o}hringer}, H. 2009,
  \aap, 498, 361

\bibitem[{{Reiprich} \& {B{\"o}hringer}(2002)}]{Reiprich02}
{Reiprich}, T.~H. \& {B{\"o}hringer}, H. 2002, \apj, 567, 716

\bibitem[{{Romer} {et~al.}(2000){Romer}, {Nichol}, {Holden}, {Ulmer}, {Pildis},
  {Merrelli}, {Adami}, {Burke}, {Collins}, {Metevier}, {Kron}, \&
  {Commons}}]{Romer00}
{Romer}, A.~K., {Nichol}, R.~C., {Holden}, B.~P., {et~al.} 2000, \apjs, 126,
  209

\bibitem[{{Romer} {et~al.}(2001){Romer}, {Viana}, {Liddle}, \&
  {Mann}}]{Romer01}
{Romer}, A.~K., {Viana}, P.~T.~P., {Liddle}, A.~R., \& {Mann}, R.~G. 2001,
  \apj, 547, 594

\bibitem[{{Rosati} {et~al.}(2002){Rosati}, {Borgani}, \& {Norman}}]{Rosati02}
{Rosati}, P., {Borgani}, S., \& {Norman}, C. 2002, \araa, 40, 539

\bibitem[{{Rykoff} {et~al.}(2008){Rykoff}, {McKay}, {Becker}, {Evrard},
  {Johnston}, {Koester}, {Rozo}, {Sheldon}, \& {Wechsler}}]{Rykoff08}
{Rykoff}, E.~S., {McKay}, T.~A., {Becker}, M.~R., {et~al.} 2008, \apj, 675,
  1106

\bibitem[{{Schuecker} {et~al.}(2003){Schuecker}, {B{\"o}hringer}, {Collins}, \&
  {Guzzo}}]{Schuecker03}
{Schuecker}, P., {B{\"o}hringer}, H., {Collins}, C.~A., \& {Guzzo}, L. 2003,
  \aap, 398, 867

\bibitem[{{Schuecker} {et~al.}(2004){Schuecker}, {B{\"o}hringer}, \&
  {Voges}}]{Schuecker04}
{Schuecker}, P., {B{\"o}hringer}, H., \& {Voges}, W. 2004, \aap, 420, 61

\bibitem[{{Sehgal} {et~al.}(2008){Sehgal}, {Hughes}, {Wittman}, {Margoniner},
  {Tyson}, {Gee}, \& {dell'Antonio}}]{Sehgal08}
{Sehgal}, N., {Hughes}, J.~P., {Wittman}, D., {et~al.} 2008, \apj, 673, 163

\bibitem[{{Sunyaev} \& {Zeldovich}(1980)}]{Sunyaev80}
{Sunyaev}, R.~A. \& {Zeldovich}, I.~B. 1980, \araa, 18, 537

\bibitem[{{Sunyaev} \& {Zeldovich}(1972)}]{Sunyaev72}
{Sunyaev}, R.~A. \& {Zeldovich}, Y.~B. 1972, Comments on Astrophysics and Space
  Physics, 4, 173

\bibitem[{{Szabo} {et~al.}(2011){Szabo}, {Pierpaoli}, {Dong}, {Pipino}, \&
  {Gunn}}]{Szabo11}
{Szabo}, T., {Pierpaoli}, E., {Dong}, F., {Pipino}, A., \& {Gunn}, J. 2011,
  \apj, 736, 21

\bibitem[{{{\v S}uhada} {et~al.}(2010){{\v S}uhada}, {Song}, {B{\"o}hringer},
  {Benson}, {Mohr}, {Fassbender}, {Finoguenov}, {Pierini}, {Pratt},
  {Andersson}, {Armstrong}, \& {Desai}}]{Suhada10}
{{\v S}uhada}, R., {Song}, J., {B{\"o}hringer}, H., {et~al.} 2010, \aap, 514,
  L3

\bibitem[{{Vanderlinde} {et~al.}(2010){Vanderlinde}, {Crawford}, {de Haan},
  {Dudley}, {Shaw}, {Ade}, {Aird}, {Benson}, {Bleem}, {Brodwin}, {Carlstrom},
  {Chang}, {Crites}, {Desai}, {Dobbs}, {Foley}, {George}, {Gladders}, {Hall},
  {Halverson}, {High}, {Holder}, {Holzapfel}, {Hrubes}, {Joy}, {Keisler},
  {Knox}, {Lee}, {Leitch}, {Loehr}, {Lueker}, {Marrone}, {McMahon}, {Mehl},
  {Meyer}, {Mohr}, {Montroy}, {Ngeow}, {Padin}, {Plagge}, {Pryke}, {Reichardt},
  {Rest}, {Ruel}, {Ruhl}, {Schaffer}, {Shirokoff}, {Song}, {Spieler},
  {Stalder}, {Staniszewski}, {Stark}, {Stubbs}, {van Engelen}, {Vieira},
  {Williamson}, {Yang}, {Zahn}, \& {Zenteno}}]{Vanderlinde10}
{Vanderlinde}, K., {Crawford}, T.~M., {de Haan}, T., {et~al.} 2010, \apj, 722,
  1180

\bibitem[{{Vikhlinin} {et~al.}(2009){Vikhlinin}, {Burenin}, {Ebeling},
  {Forman}, {Hornstrup}, {Jones}, {Kravtsov}, {Murray}, {Nagai}, {Quintana}, \&
  {Voevodkin}}]{Vikhlinin09}
{Vikhlinin}, A., {Burenin}, R.~A., {Ebeling}, H., {et~al.} 2009, \apj, 692,
  1033

\bibitem[{{Vikhlinin} {et~al.}(2006){Vikhlinin}, {Kravtsov}, {Forman}, {Jones},
  {Markevitch}, {Murray}, \& {Van Speybroeck}}]{Vikhlinin06}
{Vikhlinin}, A., {Kravtsov}, A., {Forman}, W., {et~al.} 2006, \apj, 640, 691

\bibitem[{{Vikhlinin} {et~al.}(1998){Vikhlinin}, {McNamara}, {Forman}, {Jones},
  {Quintana}, \& {Hornstrup}}]{Vikhlinin98}
{Vikhlinin}, A., {McNamara}, B.~R., {Forman}, W., {et~al.} 1998, \apj, 502, 558

\bibitem[{{Voges} {et~al.}(1999){Voges}, {Aschenbach}, {Boller},
  {Br{\"a}uninger}, {Briel}, {Burkert}, {Dennerl}, {Englhauser}, {Gruber},
  {Haberl}, {Hartner}, {Hasinger}, {K{\"u}rster}, {Pfeffermann}, {Pietsch},
  {Predehl}, {Rosso}, {Schmitt}, {Tr{\"u}mper}, \& {Zimmermann}}]{Voges99}
{Voges}, W., {Aschenbach}, B., {Boller}, T., {et~al.} 1999, \aap, 349, 389

\bibitem[{{Watson} {et~al.}(2009){Watson}, {Schr{\"o}der}, {Fyfe}, {Page},
  {Lamer}, {Mateos}, {Pye}, {Sakano}, {Rosen}, {Ballet}, {Barcons}, {Barret},
  {Boller}, {Brunner}, {Brusa}, {Caccianiga}, {Carrera}, {Ceballos}, {Della
  Ceca}, {Denby}, {Denkinson}, {Dupuy}, {Farrell}, {Fraschetti}, {Freyberg},
  {Guillout}, {Hambaryan}, {Maccacaro}, {Mathiesen}, {McMahon}, {Michel},
  {Motch}, {Osborne}, {Page}, {Pakull}, {Pietsch}, {Saxton}, {Schwope},
  {Severgnini}, {Simpson}, {Sironi}, {Stewart}, {Stewart}, {Stobbart}, {Tedds},
  {Warwick}, {Webb}, {West}, {Worrall}, \& {Yuan}}]{Watson09}
{Watson}, M.~G., {Schr{\"o}der}, A.~C., {Fyfe}, D., {et~al.} 2009, \aap, 493,
  339

\bibitem[{{Wen} {et~al.}(2009){Wen}, {Han}, \& {Liu}}]{Wen09}
{Wen}, Z.~L., {Han}, J.~L., \& {Liu}, F.~S. 2009, \apjs, 183, 197

\bibitem[{{Wilms} {et~al.}(2000){Wilms}, {Allen}, \& {McCray}}]{Wilms00}
{Wilms}, J., {Allen}, A., \& {McCray}, R. 2000, \apj, 542, 914

\bibitem[{{Yu} {et~al.}(2011){Yu}, {Tozzi}, {Borgani}, {Rosati}, \&
  {Zhu}}]{Yu11}
{Yu}, H., {Tozzi}, P., {Borgani}, S., {Rosati}, P., \& {Zhu}, Z.-H. 2011, \aap,
  529, A65

\end{thebibliography}

\clearpage
\begin{landscape}
\begin{table}
\caption{\label{} The first 25 entries of the first cluster sample}
{\tiny
\begin{tabular}{|c|l|r|r|c|r|c|r|r|r|r|r|r|r|r|r|r|r|}
\hline
  \multicolumn{1}{|c|}{detid\tablefootmark{a}} &
  \multicolumn{1}{c|}{Name\tablefootmark{a}} &
  \multicolumn{1}{c|}{ra\tablefootmark{a}} &
  \multicolumn{1}{c|}{dec\tablefootmark{a}} &
  \multicolumn{1}{c|}{obsid\tablefootmark{a}} &
  \multicolumn{1}{c|}{z\tablefootmark{b}} &
  \multicolumn{1}{c|}{scale} &
  \multicolumn{1}{c|}{$R_{ap}$} &
  \multicolumn{1}{c|}{$R_{500}$} &
  \multicolumn{1}{c|}{$T_{ap}$} &
  \multicolumn{1}{c|}{$+ eT_{ap}$} &
  \multicolumn{1}{c|}{$- eT_{ap}$} &
  \multicolumn{1}{c|}{$F_{ap}$\tablefootmark{c}} &
  \multicolumn{1}{c|}{$+ eF_{ap}$} &
  \multicolumn{1}{c|}{$- eF_{ap}$} &
  \multicolumn{1}{c|}{$L_{ap}$\tablefootmark{d}} &
  \multicolumn{1}{c|}{$+ eL_{ap}$} &
  \multicolumn{1}{c|}{$- L_{ap}$} \\

  \multicolumn{1}{|c|}{} &
  \multicolumn{1}{c|}{IAUNAME} &
  \multicolumn{1}{c|}{(deg)} &
  \multicolumn{1}{c|}{(deg)} &
  \multicolumn{1}{c|}{} &
  \multicolumn{1}{c|}{} &
  \multicolumn{1}{c|}{kpc/$''$} &
  \multicolumn{1}{c|}{(kpc)} &
  \multicolumn{1}{c|}{(kpc)} &
  \multicolumn{1}{c|}{(keV)} &
  \multicolumn{1}{c|}{(keV)} &
  \multicolumn{1}{c|}{(keV)} &
  \multicolumn{1}{c|}{} &
  \multicolumn{1}{c|}{} &
  \multicolumn{1}{c|}{} &
  \multicolumn{1}{c|}{} &
  \multicolumn{1}{c|}{} &
  \multicolumn{1}{c|}{} \\

(1)  &  (2)  &  (3)  & (4)  &   (5)   &  (6)  &   (7)   &   (8)   &  (9)   &  (10)    &  (11) &  (12) & (13) & (14) & (15) &  (16)  & (17)  &  (18)   \\
\hline 

005825 &     2XMM J003917.9+004200 &   9.82489 &   0.70013 & 0203690101 & 0.2801 & 4.24 &  89.14 &  474.51 & 1.07 & 0.33 & 0.42 &   0.51 & 0.17 & 0.22 &   1.27 &  0.46 &  0.40 \\
005901 &     2XMM J003942.2+004533 &   9.92584 &   0.75919 & 0203690101 & 0.4152 & 5.50 & 247.28 &  566.70 & 1.83 & 0.42 & 0.18 &   2.19 & 0.23 & 0.19 &  12.70 &  1.32 &  1.07 \\
006920 &     2XMM J004231.2+005114 &  10.63008 &   0.85401 & 0090070201 & 0.1468 & 2.57 & 107.87 &  464.68 & 1.41 & 0.28 & 0.33 &   1.09 & 0.41 & 0.30 &   0.62 &  0.19 &  0.17 \\
007340 &     2XMM J004252.6+004259 &  10.71952 &   0.71650 & 0090070201 & 0.2595 & 4.02 & 385.77 &  535.23 & 1.98 & 0.64 & 0.32 &   3.29 & 0.40 & 0.27 &   6.39 &  0.71 &  0.55 \\
007362 &     2XMM J004253.7-093423 &  10.72397 &  -9.57311 & 0065140201 & 0.4054 & 5.42 & 259.99 &  553.50 & 1.49 & 0.52 & 0.18 &   2.25 & 0.58 & 0.24 &  12.64 &  2.86 &  2.43 \\
007881 &     2XMM J004334.0+010106 &  10.89197 &   1.01844 & 0090070201 & 0.1741 & 2.95 & 203.84 &  549.63 & 1.34 & 0.15 & 0.13 &   4.97 & 0.27 & 0.46 &   4.15 &  0.44 &  0.34 \\
008026 &     2XMM J004350.7+004733 &  10.96148 &   0.79263 & 0090070201 & 0.4754 & 5.94 & 445.43 &  576.17 & 2.32 & 0.64 & 0.47 &   2.93 & 0.23 & 0.24 &  22.38 &  1.37 &  1.94 \\
008084 &     2XMM J004401.3+000647 &  11.00567 &   0.11323 & 0303562201 & 0.2185 & 3.53 & 307.39 &  622.06 & 1.83 & 0.40 & 0.20 &   8.71 & 1.28 & 0.83 &  11.73 &  1.69 &  1.58 \\
021508 &     2XMM J015917.2+003011 &  29.82169 &   0.50328 & 0101640201 & 0.2882 & 4.33 & 259.91 &  838.95 & 1.98 & 0.30 & 0.24 &  32.62 & 2.76 & 2.07 &  79.81 &  6.20 &  6.81 \\
021850 &     2XMM J020342.0-074652 &  30.92533 &  -7.78128 & 0411980201 & 0.4398 & 5.68 & 341.05 &  752.50 & 3.71 & 1.24 & 0.85 &  10.50 & 0.75 & 0.82 &  62.67 &  5.59 &  6.38 \\
030611 &     2XMM J023150.5-072836 &  37.96059 &  -7.47683 & 0200730401 & 0.1791 & 3.02 & 208.60 &  406.59 & 0.65 & 0.16 & 0.08 &   0.95 & 0.06 & 0.09 &   0.88 &  0.06 &  0.07 \\
030746 &     2XMM J023346.9-085054 &  38.44543 &  -8.84844 & 0150470601 & 0.2799 & 4.24 & 254.58 &  561.31 & 1.78 & 0.55 & 0.45 &   3.26 & 0.74 & 0.55 &   7.49 &  1.57 &  1.13 \\
034903 &     2XMM J030637.1-001803 &  46.65469 &  -0.30096 & 0201120101 & 0.4576 & 5.81 & 331.40 &  531.81 & 2.05 & 0.83 & 0.45 &   1.73 & 0.24 & 0.18 &  11.24 &  1.49 &  1.25 \\
042730 &     2XMM J033757.5+002900 &  54.48959 &   0.48351 & 0036540101 & 0.3232 & 4.69 & 253.05 &  566.44 & 1.80 & 0.34 & 0.37 &   3.00 & 0.40 & 0.33 &   9.22 &  0.93 &  1.00 \\
080229 &     2XMM J073605.9+433906 & 114.02470 &  43.65179 & 0083000101 & 0.4282 & 5.60 & 386.12 &  752.33 & 3.87 & 0.65 & 0.46 &  11.22 & 0.65 & 0.47 &  58.15 &  3.49 &  2.67 \\
083366 &     2XMM J075121.7+181600 & 117.84073 &  18.26679 & 0111100301 & 0.3882 & 5.28 & 316.50 &  501.14 & 1.20 & 0.30 & 0.20 &   1.65 & 0.28 & 0.25 &   8.44 &  1.27 &  1.55 \\
083482 &     2XMM J075427.4+220949 & 118.61452 &  22.16371 & 0110070401 & 0.3969 & 5.35 & 304.78 &  643.87 & 2.18 & 0.55 & 0.43 &   5.70 & 0.70 & 0.58 &  26.81 &  4.35 &  3.06 \\
089735 &     2XMM J082746.9+263508 & 126.94574 &  26.58556 & 0103260601 & 0.3869 & 5.26 & 236.90 &  530.40 & 1.69 & 1.05 & 0.46 &   1.63 & 0.55 & 0.33 &   7.93 &  2.19 &  2.51 \\
089885 &     2XMM J083146.1+525056 & 127.94516 &  52.84719 & 0092800201 & 0.5383 & 6.34 & 513.76 &  565.42 & 3.47 & 1.08 & 0.72 &   2.26 & 0.09 & 0.09 &  20.16 &  1.23 &  0.56 \\
090256 &     2XMM J083454.8+553422 & 128.72859 &  55.57287 & 0143653901 & 0.2421 & 3.82 & 286.38 & 1102.80 & 3.40 & 0.25 & 0.24 & 165.21 & 3.59 & 4.71 & 258.02 &  5.58 &  6.62 \\
090966 &     2XMM J083724.7+553249 & 129.35324 &  55.54712 & 0143653901 & 0.2767 & 4.21 & 315.60 &  677.20 & 1.83 & 0.43 & 0.24 &  11.90 & 1.67 & 1.61 &  26.22 &  4.04 &  2.91 \\
092117 &     2XMM J084701.9+345114 & 131.75794 &  34.85384 & 0107860501 & 0.4643 & 5.86 & 369.31 &  582.95 & 2.73 & 1.20 & 0.63 &   2.72 & 0.16 & 0.17 &  18.71 &  1.01 &  1.04 \\
092718 &     2XMM J084847.8+445611 & 132.19968 &  44.93637 & 0085150101 & 0.5753 & 6.55 & 353.92 &  567.36 & 1.92 & 0.38 & 0.17 &   2.53 & 0.20 & 0.14 &  30.39 &  3.25 &  1.00 \\
097911 &     2XMM J092545.5+305858 & 141.43996 &  30.98303 & 0200730101 & 0.5865 & 6.61 & 357.18 &  713.02 & 3.57 & 0.75 & 0.70 &   7.55 & 0.51 & 0.48 &  86.75 &  6.62 &  7.56 \\
098728 &     2XMM J093205.0+473320 & 143.02097 &  47.55565 & 0203050701 & 0.2248 & 3.61 & 259.96 &  567.30 & 1.36 & 0.22 & 0.23 &   5.05 & 0.77 & 0.80 &   7.46 &  1.32 &  1.05 \\

\hline
\end{tabular}
}
\tablefoot{ The full catalogue is available at CDS and contains the information given in Columns (1)-(33) in Table 2. 
 In addition, the cluster catalogue with the X-ray-optical overlay and fitted spectra 
for each cluster is publicly available from http://www.aip.de/groups/xray/XMM$_{-}$SDSS$_{-}$CLUSTERS. \\ 
\tablefoottext{a}{All these parameters are extracted from the 2XMMi-DR3 catalogue.} \\
\tablefoottext{b}{The cluster redshift from col. (27) or col. (28).}\\     
\tablefoottext{c}{Aperture X-ray flux $F_{ap}$ [0.5-2] keV and its positive and negative errors in units of $10^{-14}$\ erg\ cm$^{-2}$\ s$^{-1}$.}\\
\tablefoottext{d}{Aperture X-ray luminosity $L_{ap}$ [0.5-2] keV and its positive and negative errors in units of $10^{42}$\ erg\ s$^{-1}$.} \\
\tablefoottext{e}{X-ray bolometric luminosity $L_{500}$ and its error in units of $10^{42}$\ erg\ s$^{-1}$.} \\
\tablefoottext{f}{The cluster mass $M_{500}$ and its error  in units of $10^{13}$\ M$\odot$.}\\
\tablefoottext{g}{The Galactic HI column in units $10^{22}$\ cm$^{-2}$.}\\
\tablefoottext{h}{These parameters are obtained from first catalogue in col. (31).} \\
\tablefoottext{i}{These parameters are extracted from SDSS-DR8 data.}  \\
\tablefoottext{j}{The names of the optical catalogues which detected the cluster.} \\
\tablefoottext{k}{The known X-ray cluster names from NED or SIMBAD.} \\  
   }
\end{table}
\end{landscape}

\begin{landscape}
\addtocounter{table}{-1}
\begin{table}
\caption{\label{} The first 25 entries of the first cluster sample, continued.}
{\tiny
\begin{tabular}{|c|r|r|r|r|r|l|r|r|r|r|r|r|l|l|r|}
\hline
  \multicolumn{1}{|c|}{detid\tablefootmark{a}} &
  \multicolumn{1}{c|}{$L_{500}$\tablefootmark{e}} &
  \multicolumn{1}{c|}{$\pm eL_{500}$} &
  \multicolumn{1}{c|}{$M_{500}$\tablefootmark{f}} &
  \multicolumn{1}{c|}{$\pm eM_{500}$} &
  \multicolumn{1}{c|}{nH\tablefootmark{g}} &
  \multicolumn{1}{c|}{objid\tablefootmark{h}} &
  \multicolumn{1}{c|}{RA\tablefootmark{h}} &
  \multicolumn{1}{c|}{DEC\tablefootmark{h}} &
  \multicolumn{1}{c|}{$z_{p}$\tablefootmark{h}} &
  \multicolumn{1}{c|}{$z_{s}$\tablefootmark{i}} &
  \multicolumn{1}{c|}{$N_{z_{s}}$\tablefootmark{i}} &
  \multicolumn{1}{c|}{offset} &
  \multicolumn{1}{c|}{opt-cats\tablefootmark{j}} &
  \multicolumn{1}{c|}{known X-ray\tablefootmark{k}} &
  \multicolumn{1}{c|}{ref.} \\

  \multicolumn{1}{|c|}{} &
  \multicolumn{1}{c|}{} &
  \multicolumn{1}{c|}{} &
  \multicolumn{1}{c|}{} &
  \multicolumn{1}{c|}{} &
  \multicolumn{1}{c|}{} &
  \multicolumn{1}{c|}{} &
  \multicolumn{1}{c|}{} &
  \multicolumn{1}{c|}{} &
  \multicolumn{1}{c|}{} &
  \multicolumn{1}{c|}{} &
  \multicolumn{1}{c|}{} &
  \multicolumn{1}{c|}{(kpc)} &
  \multicolumn{1}{c|}{} &
  \multicolumn{1}{c|}{CLG} &
  \multicolumn{1}{c|}{} \\

  (1)  &  (19)  & (20)  &  (21)  &  (22)   &  (23)    &  (24)  &   (25)   &  (26) & (27) & (28)  & (29)& (30)& (31)& (32)& (33)  \\
\hline

005825 &    6.93 &   1.96 &  4.05 &  1.03 & 0.0198 & J003916.6+004215     &   9.82500 &   0.69981 & 0.2942 & 0.2801 &  1 &   5.10 & WHL,maxBCG               & -                              &  - \\
005901 &   34.56 &   3.06 &  8.04 &  1.62 & 0.0195 & 587731187282018514   &   9.92728 &   0.76163 & 0.3873 & 0.4152 &  4 &  56.11 & GMBCG,WHL,AMF            & -                              &  - \\
006920 &    3.82 &   0.76 &  3.29 &  0.79 & 0.0179 & 588015510347776148   &  10.63096 &   0.85021 & 0.1532 & 0.1468 &  6 &  36.04 & GMBCG,maxBCG             & -                              &  - \\
007340 &   13.64 &   1.53 &  5.68 &  1.20 & 0.0178 & J004252.7+004306     &  10.71964 &   0.71845 & 0.2676 & 0.2595 &  5 &  28.19 & WHL                      & [PBG2005] 03                   &  1 \\
007362 &   28.75 &   6.27 &  7.40 &  1.66 & 0.0270 & 587727226768523625   &  10.72131 &  -9.57365 & 0.4054 & 0.0000 &  0 &  52.21 & GMBCG,WHL                & -                              &  - \\
007881 &   11.95 &   0.89 &  5.61 &  1.17 & 0.0182 & 588015510347907142   &  10.89606 &   1.01972 & 0.1661 & 0.1741 &  7 &  45.59 & GMBCG,WHL,AMF            & [PBG2005] 1                    &  1 \\
008026 &   48.17 &   4.51 &  9.06 &  1.82 & 0.0179 & 587731187282477303   &  10.95832 &   0.78838 & 0.4929 & 0.4754 &  1 & 113.06 & GMBCG,WHL                & -                              &  - \\
008084 &   30.15 &   4.29 &  8.52 &  1.77 & 0.0168 & 588015509274165366   &  11.00568 &   0.11512 & 0.2580 & 0.2185 &  3 &  24.13 & GMBCG,WHL                & -                              &  - \\
021508 &  249.85 &  29.20 & 22.56 &  4.45 & 0.0234 & 588015509819293880   &  29.82170 &   0.50343 & 0.2882 & 0.0000 &  0 &   2.25 & GMBCG                    & [VMF98] 021                    &  2 \\
021850 &  222.60 &  29.56 & 19.37 &  3.86 & 0.0187 & 587727884967346553   &  30.92570 &  -7.78196 & 0.4834 & 0.4398 &  1 &  16.02 & GMBCG,AMF                & -                              &  - \\
030611 &    1.85 &   0.11 &  2.28 &  0.53 & 0.0318 & J023149.7-072834     &  37.95694 &  -7.47613 & 0.2039 & 0.1791 &  2 &  40.11 & WHL                      & -                              &  - \\
030746 &   19.75 &   3.68 &  6.70 &  1.47 & 0.0300 & 587724240688316594   &  38.44672 &  -8.84924 & 0.2799 & 0.0000 &  0 &  23.01 & GMBCG,WHL,maxBCG,AMF     & -                              &  - \\
034903 &   27.30 &   3.53 &  6.98 &  1.46 & 0.0627 & 588015508752892483   &  46.65299 &  -0.30216 & 0.4120 & 0.4576 &  1 &  43.63 & GMBCG                    & -                              &  - \\
042730 &   24.46 &   2.52 &  7.22 &  1.48 & 0.0614 & 588015509830107502   &  54.48798 &   0.48583 & 0.3220 & 0.3232 &  5 &  47.71 & GMBCG,WHL                & [PBG2005] 13                   &  1 \\
080229 &  212.68 &  16.29 & 19.09 &  3.67 & 0.0549 & J073605.8+433908     & 114.02408 &  43.65224 & 0.4011 & 0.4282 &  2 &  12.85 & WHL                      & -                              &  - \\
083366 &   14.50 &   2.56 &  5.39 &  1.20 & 0.0520 & 8718.0.0.8718        & 117.83460 &  18.26690 & 0.3969 & 0.3882 &  1 & 110.47 & AMF                      & -                              &  - \\
083482 &   71.54 &  10.89 & 11.54 &  2.37 & 0.0617 & J075427.2+220941     & 118.61320 &  22.16150 & 0.3937 & 0.3969 &  1 &  48.61 & WHL                      & -                              &  - \\
089735 &   20.56 &   6.14 &  6.38 &  1.58 & 0.0356 & 588016841246441787   & 126.94684 &  26.58608 & 0.4225 & 0.3869 &  2 &  21.17 & GMBCG,WHL,AMF            & -                              &  - \\
089885 &   54.55 &   3.55 &  9.23 &  1.82 & 0.0411 & J083146.4+525057     & 127.94344 &  52.84933 & 0.5383 & 0.0000 &  0 &  54.34 & WHL                      & -                              &  - \\
090256 & 1167.06 & 156.57 & 48.71 & 10.09 & 0.0429 & 587737808499048612   & 128.72875 &  55.57253 & 0.2033 & 0.2421 &  2 &   4.84 & GMBCG                    & 4C +55.16                      &  3 \\
090966 &   62.97 &   8.97 & 11.72 &  2.39 & 0.0405 & 8222                 & 129.35330 &  55.54786 & 0.2780 & 0.2767 &  2 &  11.21 & maxBCG                   & -                              &  - \\
092117 &   49.68 &   3.40 &  9.27 &  1.83 & 0.0292 & 587732482731737444   & 131.75750 &  34.85367 & 0.4725 & 0.4643 &  1 &   8.47 & GMBCG,WHL                & -                              &  - \\
092718 &   64.29 &   5.19 &  9.74 &  1.93 & 0.0279 & 23315.0.0.23315      & 132.20100 &  44.93770 & 0.5753 & 0.0000 &  0 &  38.44 & AMF                      & [VMF98] 060                    &  2 \\
097911 &  279.44 &  31.21 & 19.60 &  3.85 & 0.0178 & 18874.0.0.18874      & 141.43440 &  30.98410 & 0.5865 & 0.0000 &  0 & 116.35 & AMF                      & -                              &  - \\
098728 &   17.34 &   2.73 &  6.51 &  1.40 & 0.0129 & 1896                 & 143.01990 &  47.55527 & 0.2350 & 0.2248 &  1 &  10.65 & maxBCG,AMF               & -                              &  - \\

\hline
\end{tabular}
}
\tablebib{1- \cite{Plionis05}; 2- \cite{Vikhlinin98}; 3- \cite{Cavagnolo08}; 4- \cite{Finoguenov07}; 5-\cite{Horner08}; \\
 6-\cite{Dietrich07}; 7- \cite{Burenin07}; 8- \cite{Romer00}; 9- \cite{Basilakos04}; 10- \cite{Schuecker04}; 11- \cite{Sehgal08}
}
\end{table}
\end{landscape}


\clearpage 
\appendix

\section{Gallery}
We present a gallery of four galaxy clusters from the first cluster 
sample with different X-ray fluxes and data quality at different redshifts covering 
the whole redshift range of the sample. For each cluster, 
X-ray flux contours (0.2-4.5 keV) are overlaid on combined image from 
$r$, $i$, and $z$- SDSS images. The upper panel in each figure shows the 
X-ray-optical overlays. The field of view is $4'\times 4'$ centred on the 
X-ray cluster position. In each overlay, the cross-hair indicates the position 
of the brightest cluster galaxy (BCG), while in Figure A.4 the cross-hair 
indicates the cluster stellar mass centre although but it is obvious that the BCG is 
located at the X-ray emission peak. In each figure, the bottom panel shows 
the X-ray spectra (EPIC PN (black), MOS1 (green), MOS2 (red)) and the best 
fitting MEKAL model. The full gallery of the first cluster sample is available at
http://www.aip.de/groups/xray/XMM$_{-}$SDSS$_{-}$CLUSTERS.

\begin{figure}
  \vspace{1 cm}
    \resizebox{\hsize}{!}{\includegraphics{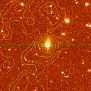}}\\ \\ 
    \resizebox{\hsize}{!}{\includegraphics[angle=-90, viewport=30  0  520 660 ,clip]{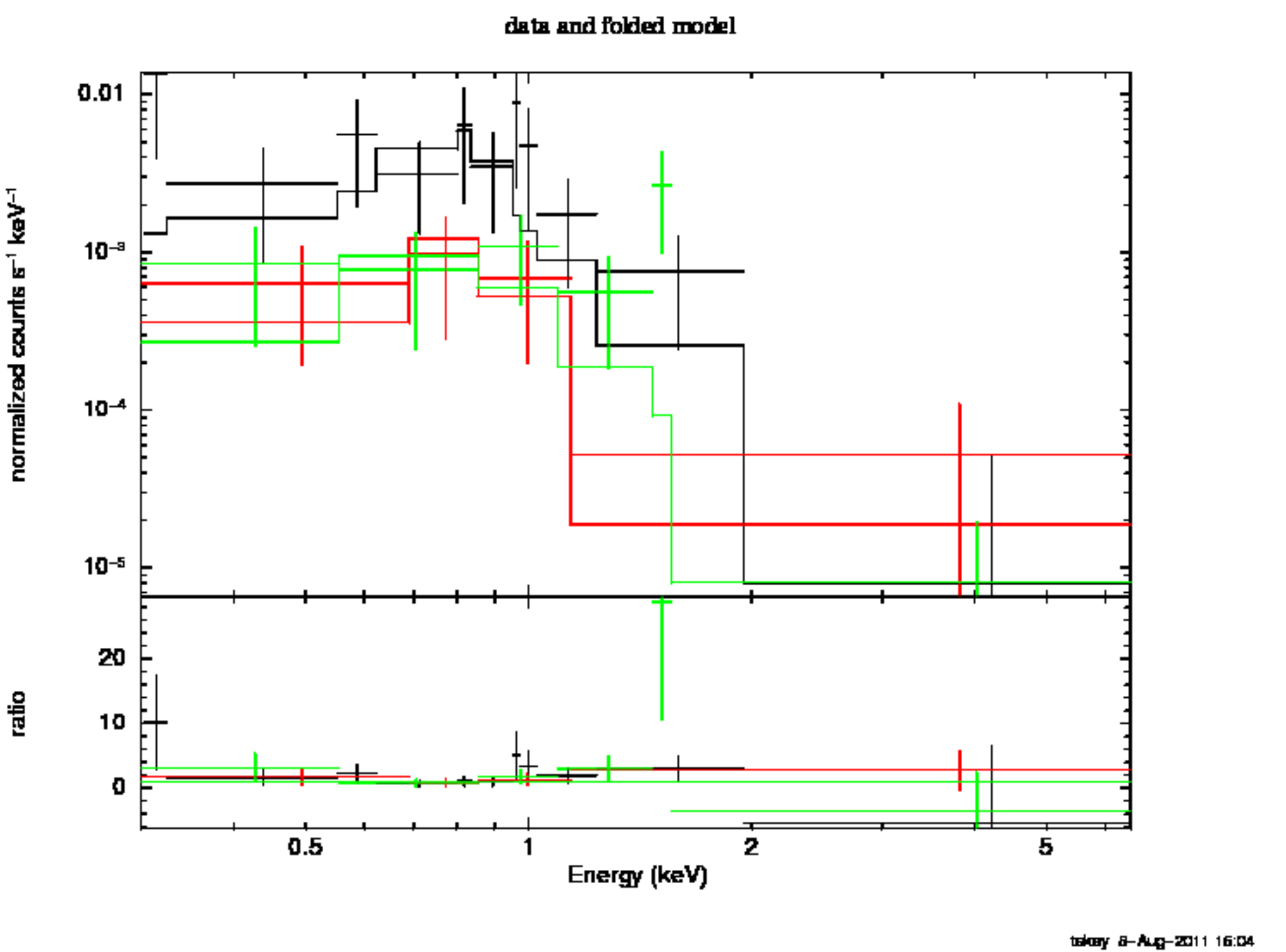}}\\ \\
    \caption{ detid = 275409: 2XMMI J143929.0+024605 at $z_{s}$ = 0.1447
   ($F_{ap}\ [0.5-2]\ keV = 0.63 \times 10^{-14}$ \ erg\ cm$^{-2}$\ s$^{-1}$).}
  \label{Fig.A.1}
\end{figure}

\begin{figure}
\vspace{1 cm}
  \resizebox{\hsize}{!}{\includegraphics{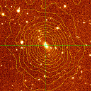}}\\ \\
  \resizebox{\hsize}{!}{\includegraphics[angle=-90, viewport=30  0  520 660 ,clip]{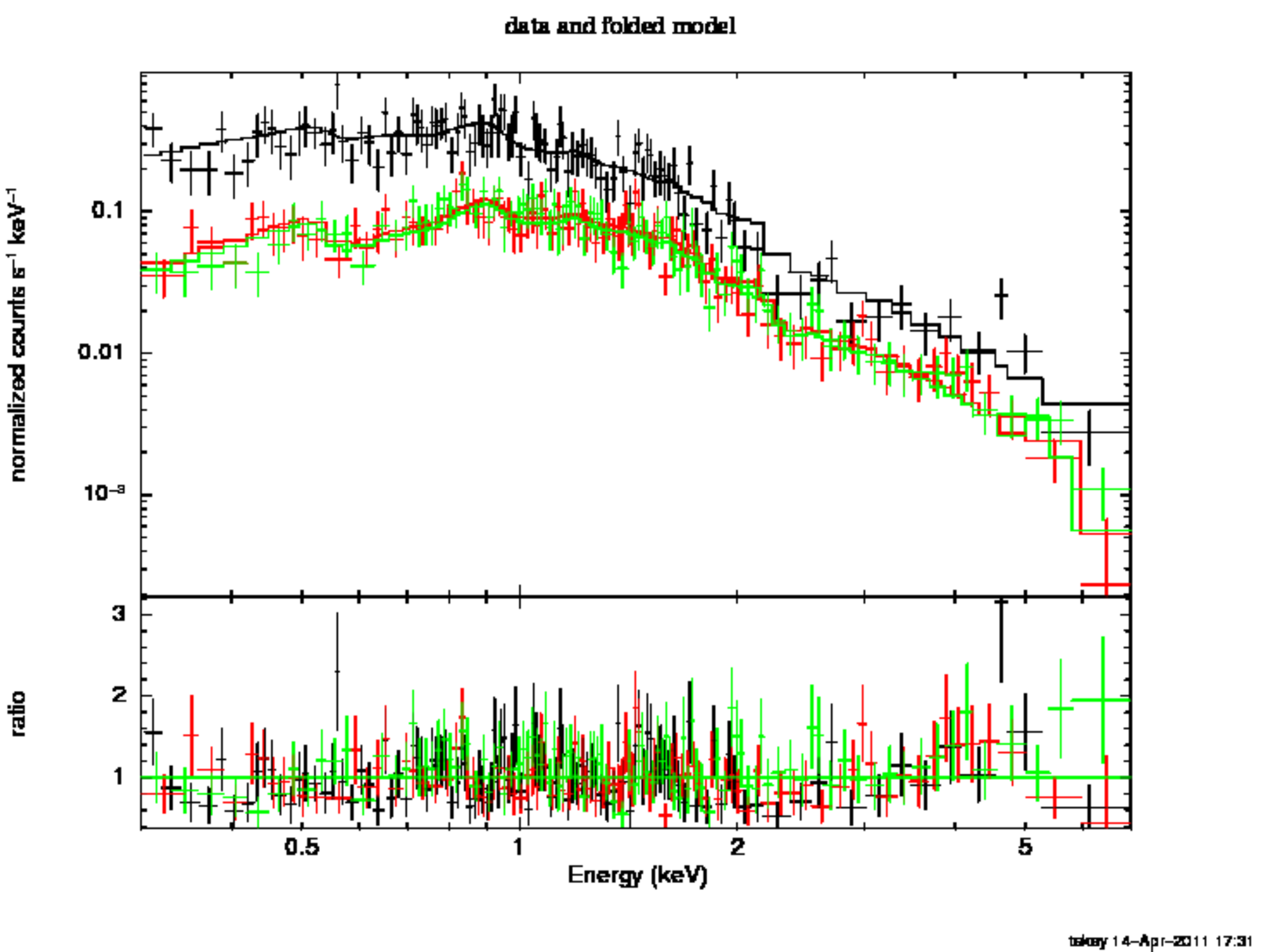}}\\ \\
  \caption{detid = 090256: 2XMM J083454.8+553422 at $z_{s}$ = 0.2421 
($F_{ap}\ [0.5-2]\ keV = 165.21 \times 10^{-14}$ \ erg\ cm$^{-2}$\ s$^{-1}$).  }
  \label{Fig.A.2}
\end{figure}

\begin{figure}
\vspace{1 cm}
  \resizebox{\hsize}{!}{\includegraphics{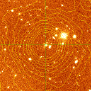}}\\ \\
  \resizebox{\hsize}{!}{\includegraphics[angle=-90, viewport=30  0  520 660 ,clip]{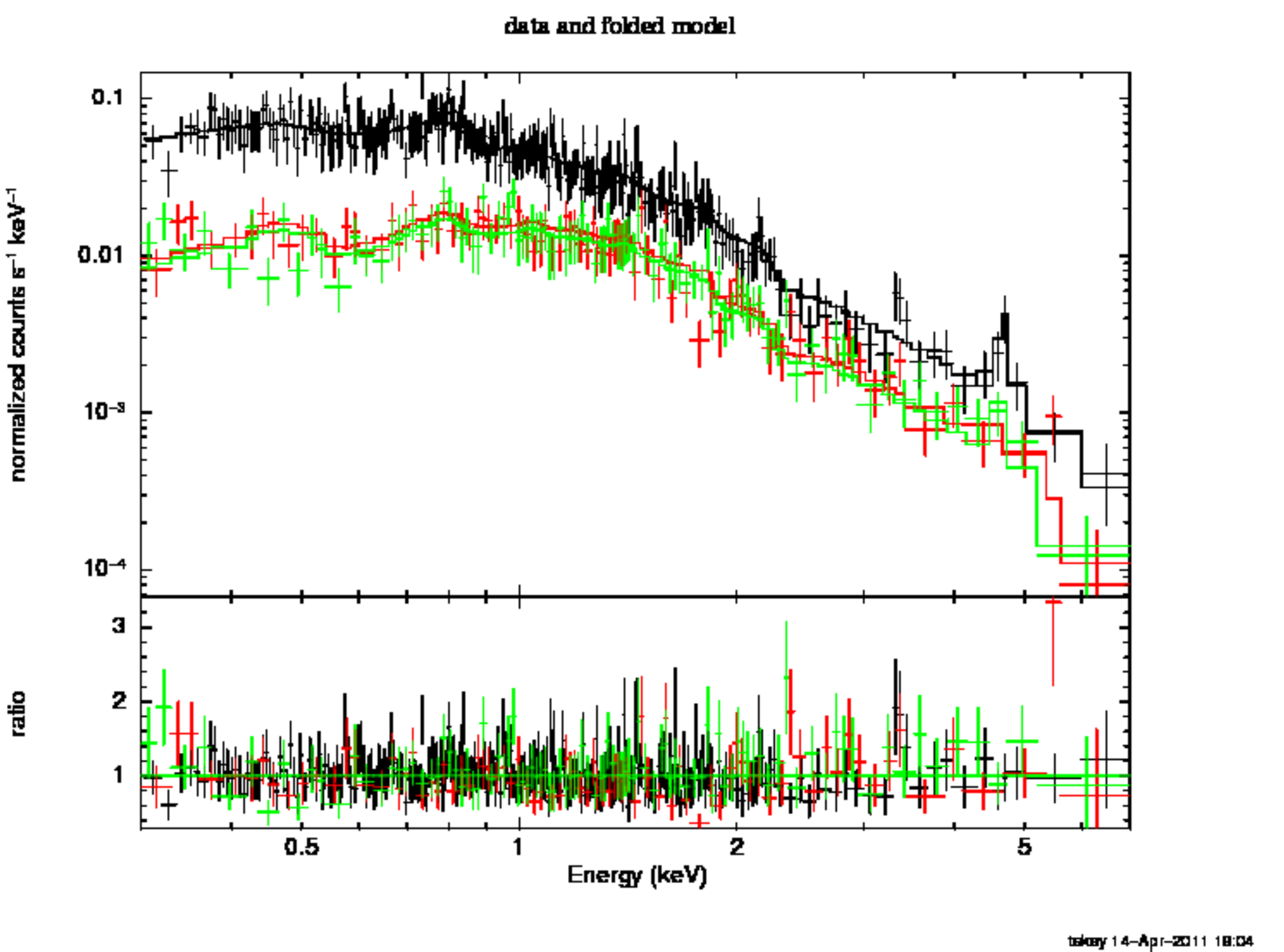}}\\ \\
  \caption{detid = 312615: 2XMM J091935.0+303157 at $z_{s}$ = 0.4273
($F_{ap}\ [0.5-2]\ keV = 16.03 \times 10^{-14}$ \ erg\ cm$^{-2}$\ s$^{-1}$).  }
  \label{Fig.A.3}
\end{figure}

\begin{figure}
\vspace{1 cm}
  \resizebox{\hsize}{!}{\includegraphics{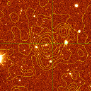}}\\ \\
  \resizebox{\hsize}{!}{\includegraphics[angle=-90, viewport=30  0  520 660 ,clip]{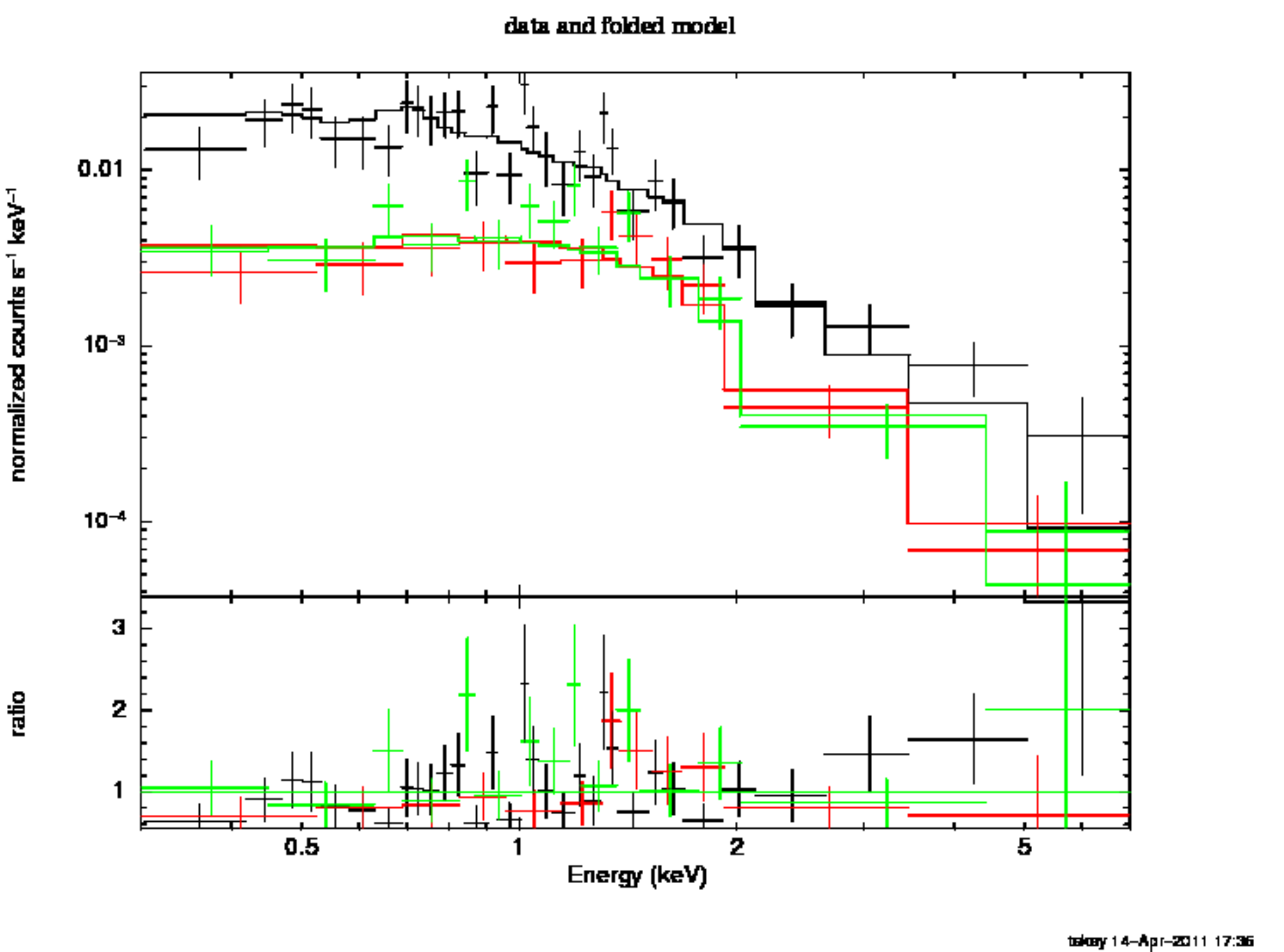}}\\ \\
  \caption{detid = 097911: 2XMM J092545.5+305858  at $z_{p}$ = 0.5865
($F_{ap}\ [0.5-2]\ keV = 7.59 \times 10^{-14}$ \ erg\ cm$^{-2}$\ s$^{-1}$). }
  \label{Fig.A.4}
\end{figure}

\end{document}